\documentclass[sn-apa]{sn-jnl}
\usepackage{graphicx}%
\usepackage{multirow}%
\usepackage{amsmath,amssymb,amsfonts,bbm}%
\usepackage{amsthm}%
\usepackage{mathrsfs}%
\usepackage[title]{appendix}%
\usepackage{xcolor}%
\usepackage{textcomp}%
\usepackage{manyfoot}%
\usepackage{booktabs}%
\usepackage{algorithm}%
\usepackage{algorithmicx}%
\usepackage{algpseudocode}%

\usepackage{listings}%
\usepackage{cleveref}
\newcommand{\Fcal}{\mathcal{F}}
\newcommand{\Hcal}{\mathcal{H}}
\newcommand{\Lcal}{\mathcal{L}}
\newcommand{\Mcal}{\mathcal{M}}
\newcommand{\Tcal}{\mathcal{T}}
\newcommand{\Scal}{\mathcal{S}}
\newcommand{\Vcal}{\mathcal{V}}
\newcommand{\R}{\mathbb{R}}
\newcommand{\E}{\mathbb{E}}

\newcommand{\one}{\mathbbm{1}}

\newcommand{\intsimplexk}{\mathring{\Scal_K}}

\newtheorem{theorem}{Theorem}
\newtheorem{proposition}[theorem]{Proposition}%
\newtheorem{remark}{Remark}%

\begin{document}
	
	\title[Onflow portfolio allocation]{Onflow: a model free, online portfolio allocation algorithm robust to transaction fees}
	
	\author{\fnm{Gabriel} \sur{Turinici}\footnote{ORCID : 0000-0003-2713-006X}}\email{Gabriel.Turinici@dauphine.fr}

	\author{\fnm{Pierre} \sur{Brugiere} \footnote{ORCID : 0000-0002-8716-9145}}\email{brugiere@ceremade.dauphine.fr }

	  \affil{\orgdiv{CEREMADE}, \orgname{Universit\'e Paris Dauphine - PSL} \\ \orgaddress{\street{Place du Marechal de Lattre de Tassigny}, \city{Paris}, \postcode{75116}, \state{Paris}, \country{FRANCE}} 
		\\ \ \\ 
		\today}

\abstract{
We introduce Onflow, a reinforcement learning method for optimizing portfolio allocation via gradient flows. Our approach dynamically adjusts portfolio allocations to maximize expected log returns while accounting for transaction costs. Using a softmax parameterization, Onflow updates allocations through an ordinary differential equation derived from gradient flow methods. 
This algorithm belongs to the large class of stochastic optimization procedures; we measure its efficiency by comparing our results to the mathematical theoretical values in a log-normal framework and to standard benchmarks from the 'old NYSE' dataset.
		
For log-normal assets with zero transaction costs, Onflow replicates Markowitz optimal portfolio, achieving the best possible allocation. Numerical experiments from the 'old NYSE' dataset show that Onflow leads to dynamic asset allocation strategies whose performances are: a) comparable to benchmark strategies such as Cover's Universal Portfolio or Helmbold  et al. ``multiplicative updates'' approach when transaction costs are zero, and b) better than previous procedures when transaction costs are high. Onflow can even remain efficient in regimes where other dynamical allocation techniques do not work anymore.
		
Onflow is a promising portfolio management strategy that relies solely on observed prices, requiring no assumptions about asset return distributions. This makes it robust against model risk, offering a practical solution for real-world trading strategies.
}
	


	
\maketitle

	\section{Motivation and literature review}\label{sec:intro}
	
	Ever since the advent of modern portfolio theory, reliable information on the statistical properties of the financial time series  has been a crucial determinant of the portfolio performance. Formulated in a mean-variance setting, the classical approach of  \cite{markovitz_portfolio_theory} promises optimal performance when the {\bf future} first and second order moments are known. In general this information is highly uncertain and in practice the quality of the result is far from the expected level. To cure this empirical drawback, several approaches have been proposed: 
	\cite{kelly1956new} analyzed optimal bet size in investment portfolios,
	\cite{black1990asset}  
	modeled the expected return as variables that are updated, by investor convictions, though a Bayesian mechanism, while \cite{cover91} introduced the Universal portfolio to profit from the long term exponential behavior and obtain results that are comparable with the best constant rebalanced portfolio chosen in hindsight. This latter approach uses no assumption whatsoever about the statistical properties of the asset time series and was followed by a large body of literature aiming to produce performances robust to variations in the model parameters. Among such follow-ups we will focus on the online learning approaches which enters the general framework of reinforcement learning, where data is fed directly into a strategy without any model in between. 
	In particular \cite{Helmbold98} proposed a first  version using multiplicative updates and a relative cross-entropy loss function, 
	\cite{li_moving_avg_portf_2015} continued along these lines assuming a reversion to the mean while \cite{blum1997universal} explored the theoretical and practical implications of transaction costs. Lastly,   \cite{kirby_low_turnover_12} proposed low turnover strategies. 
	On the other hand, \cite{borodin2003can} introduced the Anticor algorithm, which  exploits the general idea of correlation between assets. 
	For additional findings on online portfolio selection we refer to 
	the 
	reviews of \cite{li2014online_survey}, \cite{sato2019modelfree} 
	and \cite{sun_review_rl_trading2023}
	while for a more machine learning orientation see~\cite{cesa2006prediction}
	and \cite[Chapter 3]{book_fin_rl12}; finally 
	see \cite{li2016olps} for an open source toolbox to test algorithms. More recently, 
	\cite{zhang_combining_2021} proposed a strategy combining different experts,  
	\cite{he_new_2024} made available a literature review and an extension of the Anticor approach using dynamic time warping as a similarity distance; 
	\cite{deep_policy_grad_19} investigated policy gradient style deep reinforcement learning approaches and \cite{ngo_does_2023}  compared reinforcement learning and deep learning methods in portfolio optimization.
	
	Remaining in this framework of  online\footnote{Online is understood here in the sense of \cite{li2014online_survey} and designates a special procedure that makes decisions based on continuously streamed data. This is to be contrasted with usual supervised procedures that first use data to learn a model and then make predictions.}, no-hindsight, reinforcement learning, model-free approaches we present here an algorithm using the gradient flow concept instead of discrete updates that can treat in an intrinsic way the transactions costs. The portfolio allocation is parameterized through a softmax function.
	
	The outline of the paper is the following; in section~\ref{sec:notations} we introduce the Onflow algorithm; subsequently, in section~\ref{sec:theory} some theoretical results are presented. In section~\ref{sec:numerical} we test the performance of the procedure on several benchmarks from the literature and in section~\ref{sec:conclusion} we conclude with additional remarks.
	
	\section{Onflow algorithm: intuition and formal definition}\label{sec:notations}

	Consider a market $\Mcal$ containing $K$ financial assets and $T$ time instants $t\in \Tcal :=\{1,...,T\}$; $T$ can be either finite or infinite.
	We denote $S^k_t$ the value at time $t$ of the asset $k$ and assume $S^k_0=1, S^k_t >0, \forall k, t$.
	The price relatives $f_t^k$ are defined as 
	$f^k_t = S^{k}_{t} /  S^{k}_{t-1}$, $\forall t\in \Tcal$. 
	
	A portfolio is characterized by a set of $K$ weights $\pi=(\pi(1),..., \pi(K))$.
	At any time $t$ the quotient of the wealth invested in the asset $k$ with respect to total portfolio value is set to $\pi(k)$ which means that the $\pi(k)$ sum up to one. 
	We  suppose that each $\pi(k)$ is positive, i.e., no short selling is allowed. In this case $\pi$ belongs to the unit simplex $\Scal_K$ of dimension $K$~:
	\begin{equation}
		\Scal_K = \left\{w = (w_k)_{k=1}^K \in \R^K : w_k \ge0, \sum_{k=1}^K w_k = 1 \right\}.
	\end{equation}
	We will denote $\intsimplexk$
	the interior  of  $\Scal_K$ i.e.
	\begin{equation}
		\mathring{\Scal_K}
		= \left\{w = (w_k)_{k=1}^K \in \R^K : w_k > 0, \sum_{k=1}^K w_k = 1\right\}.
	\end{equation}

	When $\pi$ is constant over time, we obtain the so called Constant Rebalanced Portfolio (CRP) also called a 'Constant Mix' portfolio. Note that a CRP is a dynamic investment strategy because the price evolution may induce a shift in the proportions $\pi$ which have to be reset to the prescribed values.

	We denote $f_t$ the vector with components $f^k_t$, $k\le K$, $t\in \Tcal$. A portfolio 
	with initial value $\Vcal_0(\pi)$ at $t=0$
	and weights $\pi_t$ chosen at time $t-1$ has the value  $\Vcal_t(\pi) $ at time $t$ with~:
	\begin{equation}
		\Vcal_0(\pi)=1, \ \forall t \in \Tcal~: \Vcal_{t}(\pi) = \Vcal_0(\pi) \prod_{s=1}^{t} \langle \pi_s, f_s\rangle = \Vcal_0(\pi) e^{\sum_{s=1}^{t} \ln(\langle \pi_s, f_s\rangle)}. \label{eq:portfolio_value}
	\end{equation}
	As a side remark, note that formula \eqref{eq:portfolio_value} can also be written
	\begin{equation}
		\ln( \Vcal_{t}(\pi) ) = 	\ln( \Vcal_0(\pi) )+ \sum_{s=1}^t \ln(\langle \pi_s, f_s\rangle) = 	\ln( \Vcal_0(\pi) )+ 
		t \cdot \frac{ \sum_{s=1}^t \ln(\langle \pi_s, f_s\rangle)}{t},
	\end{equation}
	and, if we interpret $(\pi_s,f_s)$ to be samples from 
	some joint distribution denoted $(\pi,f)$
	we recognize in $ \frac{ \sum_{s=1}^t \ln(\langle \pi_s, f_s\rangle)}{t}$ an empirical estimator for $\E[ \ln(\langle \pi, f\rangle) ]$.
	
	These manipulations are useful to write the problem of maximizing the final value in the form of maximizing an average, which is the usual formalism in Reinforcement Learning frameworks.
	
	\subsection{Reinforcement learning framework} \label{sec:rl}
	
	Reinforcement learning (abbreviated 'RL' from now on, see \cite[Chapter 3]{sutton_reinforcement_2018} for a pedagogical introduction)  
	can be used in some situations when model-free approaches are necessary for problems involving repeated decisions, such as game play, robot maneuvering, autonomous car driving, etc. 
	
	For the reader already versed in reinforcement learning we provide below the transcription of our setting into the formal writing of a RL problem which involves~: 
	\begin{itemize}
		\item 	a sequence of time  instants: for us will be $\Tcal$
		\item  a state of the world at each time instant $t$: for us this will be the allocation $\pi_t$ and the portfolio value $\Vcal_t(\pi)$
		\item a set of actions to chose from at time $t$: for us this is $\Scal_K$ where $\pi_{t+1}$ belongs
		\item rewards $r_t$ obtained at each time $t$ depending on previous actions, see  below for the precise choice we make. Note that it is not necessary for the reward $r_t$ to result deterministically from the actions. 
		\item a strategy to choose the next action: the general prescription in reinforcement learning is to choose a probability law on the set of actions, i.e., a distribution in $\Scal_K$. 
	\end{itemize}
	Then the problem is formalized as~:
	\begin{equation}
		\textrm{ choose iteratively } \pi_{t+1} 
		\textrm{ to maximize the expected value of the rewards }
		\E [r_t]. \label{eq:rl_rewards}
	\end{equation}
	This formulation can be framed in the general set of `Policy Gradient' approaches, see \cite[Section 2.8 and Chapter 13]{sutton_reinforcement_2018}.
	
	\begin{remark}
		Compared with the reinforcement learning literature we consider that the rewards  are not discounted, i.e., a reward $r$ at time $s$ is worth as much as a reward $r$ at some other time. Such discounting is  often used when the quantity to optimize would be infinite but increasing sub-exponentially. 
		
		A possible choice for the rewards $r_t$ are the portfolio gains from time $t-1$ to $t$. This may not be a good idea because the increase could be exponential and even discounting may not help to make it finite. In view of the relation \eqref{eq:portfolio_value} above and in coherence with existing literature, it is more natural to look for procedures that maximize the expected value of 
		$r_t=\ln(\langle \pi_t, f_t\rangle)$; for instance \cite{Helmbold98} 
		chooses $\pi_{t+1}$ to maximize this expected value using a particular choice of multiplicative updates derived from an approximation of the relative entropy to the first order. 
		We will subscribe to the same convention but we add to $r_t$ a term to model the transactions costs as explained below.
\end{remark}
	
\subsection{The Onflow algorithm}

	We therefore look for iterative procedures that
	starting from the state of the portfolio and of the market $\Mcal$ up to time $t$
	adjusts $\pi_t$ into some $\pi_{t+1}$ 
	with better expected rewards.
	In coherence with the extensive literature on the gradient flows~\cite{jko}, it is natural to also require $\pi_{t+1}$ to be somehow close to $\pi_t$. Various ways to impose this proximity are possible, most of them exploiting the fact that $\pi$ is a discrete probability law on the set $\{1,...,K\}$ for instance \cite{Helmbold98} uses relative cross-entropy.
	We will parameterize $\Scal_K$ through the "softmax" function, denoted $S(\cdot)$ and defined by~: 
	\begin{equation}
		S:\R^K \to \Scal_K, \  H\in \R^K \mapsto S(H)=\pi \in \Scal_K, \pi(k) = \frac{e^{H_k}}{\sum_\ell e^{H_\ell}}.
		\label{eq:softmax_h_to_pi}
	\end{equation}
	
	\begin{remark}
		A limited amount of short selling can be easily accommodated by taking as portfolio allocation not $S(H)$ but $\pi_\lambda=(1+\lambda)S(H)-\lambda/K$, with $\lambda>0$ a fixed value; the entries sum up to $1$ but $\pi_\lambda$ is not always in $\Scal_K$ as it can have negative entries not exceeding $\lambda/K$. \label{rem:short}
	\end{remark}
	
	Let us denote with these new variables our reward function~:
	\begin{equation}
		F_{t}(H) := \ln(\langle S(H),f_t\rangle).
	\end{equation}
	So, $H_{t+1}$ could be chosen to maximize $F_t$ as a posterior best choice and is expected to be close to $H_t$ (one single new observation should not generate a complete change of strategy). 
	Note that one of the reasons why we expect $H_{t+1}$ to stay close to $H_t$ is because the transition from $H_t$ to $H_{t+1}$ can be costly in terms of transaction fees. 
	We will consider proportional transaction fees that charge a given, known, percentage of the amount sold or bought; note that in this transaction fee model moving an amount $X$ from one asset to the other will cost twice this percentage because both buying and selling are taxed; see also~\cite{blum1997universal} for additional discussions on the transaction fees models and for some optimizations that occur. We will not consider here such buy/sell optimization and to make things comparable with the literature we resume everything to a fee level $\xi>0$ and consider that for a portfolio of value $V$ switching from allocation $\tilde\pi$ to $\pi$ incurs a fee  $\xi V \sum_k |\tilde\pi(k) - \pi(k)|$.

	The allocation $\pi_t$ that was selected at time $t-1$ and before prices at $t$ were known will drift by itself 'overnight' because of the price evolution given by the price relatives $f_t$~;	a simple computation shows that the new allocation that takes into account the prices at time $t$ is~:
	\begin{equation}
		\pi_{t+}=\frac{\pi_t \odot f_t}{\langle \pi_t , f_t \rangle} = 
		\left(
		\frac{\pi_t(k) f_t(k)}{\sum_\ell \pi_t(\ell) f_t(\ell)} 
		\right)_{k=1}^K, \ 
		\odot = \textrm{ element-wise (Hadamard) product}. \label{eq:pitplus}
	\end{equation}
	
	Rebalancing a portfolio of total value $V$ whose allocation drifted to  $\pi_{t+}$
	to some target allocation $\pi$ will
	lower $V$ to $V- V\xi \sum_k |\pi_{t+}(k)-\pi(k)|$ i.e. will act by a multiplication with $1- \xi \sum_k |\pi_{t+}(k)-\pi(k)| 
	\simeq e^{-\xi \sum_k |\pi_{t+}(k)-\pi(k)|}$ where for the  approximation we used that $\xi$ is small compared to $1$\footnote{Of course, it is possible to not employ this approximation and use the exact relation at the cost of more complicated formulas involving the logarithmic derivative; we noticed however that in practice this has no impact on the results and stick with the simpler form.}. 
	As a technical detail, the absolute value $|\cdot|$ above is not smooth enough and may induce numerical instabilities in the computations; to avoid this we regularize it to $\sqrt{|\cdot|^2+a^2} -a$; such a function proved to be useful in many areas of machine learning, cf.~\cite{pseudo_huber_loss,turinici_radonsobolev_2021} and is sometime called "pseudo-Huber" loss. 
Replacing the nonsmooth $|\cdot|$ term with a pseudo-Huber loss offers several practical benefits in gradient flow dynamics: it ensures differentiability everywhere, including near zero, which improves numerical stability and convergence; moreover the smooth transition from quadratic behavior around zero to linear growth for large $|x|$ provides better conditioning and avoids abrupt gradient changes. In addition this formulation also maintains robustness to outliers while preventing gradient explosions, making it suitable for regularization tasks. Another benefit is that the resulting continuous-time dynamics is well-behaved and easier to integrate, which is critical for stable optimization in real-world applications.
 For numerical tests we set $a=10^{-6}$.

	Recalling that we are maximizing the expectation of the logarithm of the rewards, the transaction fees are therefore modeled as~:
	\begin{equation}
		G_t(H):=\xi \sum_{k=1}^K  \sqrt{[\Scal(H)(k) - \pi_{t+}(k)]^2+a^2}-a.
	\end{equation}
	
	With these provisions one can look for $H_{t+1}$ 
	close to $H_t$ and that minimizes  $\Fcal_t(H):=G_t(H)- F_{t}(H)$;
	we will define $H_{t+1}$ as follows: solve for $u\in [0,\tau]$ the ODE~\footnote{It is known that under suitable assumptions on the function $f$, minimizing $x \mapsto f(x)$ can be obtained by solving $x'(t) = - \nabla_x f(x(t))$.}: 
	\begin{equation}
		\Hcal(u=0)=H_{t}, \ 
		\frac{d}{d u} \Hcal(u) = - \nabla_H \Fcal_t(\Hcal(u)),
		\label{eq:defH_u_general_F_G}
	\end{equation}
	then set $H_{t+1}=\Hcal(\tau)$; $\tau>0$ is a parameter of our algorithm.
	Replacing the gradients $\nabla_H F_t$ and $\nabla_H G_t$ we obtain
	the following ODE~:
	\begin{equation}
		\begin{cases}
			\Hcal(u=0)=H_{t} \\
			\frac{d}{d u} \Hcal(u)=  \frac{\Scal(\Hcal(u)) \odot f_t }{ \langle \Scal(\Hcal(u)) ,f_t \rangle}
			-\Scal(\Hcal(u)) 
			\\
			- \xi  \left(\sum_k \frac{\Scal(\Hcal(u)) (k)-\pi_{t+}(k)}{\sqrt{(\Scal(\Hcal(u))(k)-\pi_{t+}(k))^2+a^2}} \Scal(\Hcal(u))(k)(\one_{k=b}-\Scal(\Hcal(u))(b))\right)_{b=1}^K.
		\end{cases}
		\label{eq:ode_formula}
	\end{equation}
	
	We used here the softmax derivation formula for $S(\cdot)$: 
	\begin{equation}
		\frac{\partial}{\partial H_b} S(H)(k) = S(H)(k) (\one_{k=b}-S(H)(b)).
		\label{eq:softmax_derivation_rule}
	\end{equation}
	We can now formally introduce the 'Onflow' algorithm, described in the Algorithm \ref{algo:onflow} whose pseudo-code is given below.
	\footnote{
		The $f_t$ are obtained from the relation
		$f_t = S_{t} /  S_{t-1}$, 
		where $S_t$ is the closing price at date $t$ for the asset. 
		The equation \eqref{eq:ode_formula} is used to compute 
		$H_{t+1}$ which  determines by the equation \eqref{eq:softmax_h_to_pi} the allocation $\pi_{t+1}$ until the closing price $S_{t+1}$ is known. 
		The ODE \eqref{eq:ode_formula} can be solved in milliseconds and 
		we assume, as in all previous studies, that it is possible to trade at the closing price a few milliseconds after the official close.
		
		Note for instance that currently this is {\bf guaranteed} by the Euronext exchange through the ``Trade at Last'' (TAL) mechanism from 17h35 to 17h40 that allows, during these 5 minutes, to trade at the closing price of the day $S_t$. So, we do not use any future information. 	
	}
	
	\begin{algorithm}
		\caption{Onflow portfolio allocation algorithm}
		\label{algo:onflow}
		{\bf Inputs: }  parameter  $\tau>0$, $a$ (default value $a=10^{-6}$).
		
		{\bf Outputs: } allocations $\pi_t$, $t\in \Tcal$.
		
		\begin{algorithmic}[1]
			\Procedure{}{}
			\State Set $t=1$, $H_t=0 \in \R^K$.
			\For{$t\in T$}
			\State  read $f_t$, compute $\pi_{t+}$ from \eqref{eq:pitplus}
			\State solve ODE \eqref{eq:ode_formula} and set $H_{t+1}=\Hcal(\tau)$,
			$\pi_{t+1}=S(H_{t+1})$ \label{algo:odestep}
			\State store $\pi_{t+1}$
			\EndFor \label{algo:endfor}
			\EndProcedure
		\end{algorithmic}
	\end{algorithm}
	
	\begin{remark}
		In general solving the ODE at line \ref{algo:odestep} is not difficult because the number of assets is in practice not too large ($2$ to $100$). Should this not be the case, one can try instead an explicit Euler numerical scheme with step $\tau$ which boils down to simple vectors addition.
	\end{remark}
	\begin{remark}
		For comparison, the $EG(\eta)$ algorithm of \cite{Helmbold98} use instead an update of the form~:
		\begin{equation}
			H_{t+1}=H_t + \tau \frac{f_t}{\langle \pi_t,f_t\rangle} + c_t,
		\end{equation}
		where $c_t$ is a constant with respect to $k\le K$ but that can change with time. It also corresponds to a Natural Policy Gradient (NPG) algorithm, see \cite{amari_natural_gradient_98} for the seminal work on the natural gradient and
		\cite[Lemma 15]{agarwal_theory_2021_cv_policy_grad} for its formulation in reinforcement learning under the policy gradient framework.
		Our proposal of using continuous gradient flow updates, implemented through the ODE \eqref{eq:ode_formula} instead of the discrete EG algorithm, introduces smoother and more controllable updates as it aligns with optimization theory by providing a principled continuous-time approach that ensures convergence under mild conditions. On the other hand the inherent geometry of the softmax formulation improves stability, mitigating overshooting and collapse that can occur with naive updates. Moreover, gradient flow  balances exploration and exploitation more effectively than direct parameter shifts. Finally, this approach connects naturally to mirror descent and KL-based regularization, offering a theoretically grounded framework widely adopted in reinforcement learning and probabilistic optimization, cf.  \cite{sutton_reinforcement_2018}.
		\label{rem:ComparisonWithEG}
		 \end{remark}
	
	\begin{remark}
		The price relatives $f_t$ are stochastic in nature and the maximization of the performance needs to take into account this fact. The standard way to deal with such circumstance is to use a stochastic optimization algorithm, variant of the Stochastic Gradient Descent introduced in \cite{robbins_stochastic_1951}; see \cite{sutton_reinforcement_2018} for its use in reinforcement learning in general and \cite{gabriel_turinici_convergence_2023} for a short self-contained convergence proof.
		When optimizing a general function $\Fcal(x)$
		this optimization algorithm converges even if, instead of the true gradient 
		$\nabla \Fcal(x)$
		only a non-biased version is used at each step instead; in practice, to lower the variance of the error, a
		sample average based on $B$ non-biased gradients can be used. 
		This means that instead of advancing $1$ step at the time one can advance $B$ steps and adapt formula \eqref{eq:ode_formula} to take into account a sample average of price relatives $f_t$,..., $f_{t+B}$.
		Note that the algorithm, as written above, corresponds to $B=1$.
	\end{remark}

	\section{Theoretical convergence results}\label{sec:theory}
	
	We present in this section a convergence result which shows that the Onflow algorithm will reach optimality under some special assumptions. More precisely, we will consider the continuous limit i.e., $\Tcal=\R_+$, no transaction fees ($\xi=0$)  and assume that the asset dynamic is log-normal. Of course, this is a simplification because in real life no asset dynamic is exactly log-normal.
	But, it is still reassuring that in this prototypical situation our algorithm is consistent and provides the expected solution. 
	As in recent works on the convergence of softmax-formulated reinforcement learning problems, see \cite{mei20_cv_softmax_policy,agarwal_theory_2021_cv_policy_grad}, we will work in the "true gradient" regime.
	
	We use the following notations 
	for the log-normal dynamics of the assets~: 
	\begin{equation}
		\frac{d S^k_t }{S^k_t} = \mu_k dt + \sum_{\ell=1}^K \sigma_{k\ell} d z_\ell(t),
	\end{equation}
	where $z_\ell$ are independent Brownian motions. Note that in general the drifts 
	$\mu=(\mu_k)_{k=1}^K$
	and $\sigma=(\sigma_{k\ell})_{k,\ell=1}^K$ are unknown. The 
	covariance matrix will be denoted $\Sigma = \sigma^T \sigma$.
	Since for any $T \ge 0$~:
	\begin{equation}
		\E [ \ln(\Vcal_T(\pi))] = 
		\E [ \ln(\Vcal_0(\pi))] + \int_0^T \frac{d}{dt} \E [ \ln(\Vcal_t(\pi))] dt,
	\end{equation}
	we can formulate as in~\cite{cont_time_univ_portf_jamshidian92}
	the log-optimum portfolio as the continuous maximization over $\Scal_K$ of
	$\frac{d}{dt} \E [ \ln(\Vcal_t(\pi))]$ which means that in this setting
	$F_t(H) :=\frac{d}{dt} \E [ \ln(\Vcal_t(S(H)))]$. Or, the Ito formula shows that~:
	\begin{equation}
		F_t(H)= R(S(H)),
		\textrm{ where }
		R(\pi) := \langle \mu, \pi \rangle - \frac{1}{2} \pi^T \Sigma \pi. 
		\label{eq:fed_f_continuous}
	\end{equation}
	Since $\xi=0$ from 
	equations  \eqref{eq:defH_u_general_F_G} 
	and \eqref{eq:fed_f_continuous} we obtain that the algorithm corresponds to solving the following ODE~:
	\begin{eqnarray}
		& \ & 
		\Hcal(0)=0 \in \R^K, \ 
		\frac{d}{d t} \Hcal(t) = \nabla_H R(S(\Hcal(t))) \ \forall t >0.
		\label{eq:def_algo_cont}
		\\ & \ & 
		\textrm{Output allocation at time } t~:~ \pi_t=S(\Hcal(t)).
		\label{eq:def_pit_cont}
	\end{eqnarray}
	On the other hand the optimal allocation $\pi^\star$ is the solution of the following problem~:
	\begin{equation}
		\max_{\pi \in \Scal_K} R(\pi). \label{eq:max_pi_cont}
	\end{equation}
	\begin{remark}
		The class of functionals defined by  
		$R^{\lambda}(\pi) := \langle \mu, \pi \rangle - \frac{\lambda}{2} \pi^T \Sigma \pi$ has as its minimizers 
		the class of the efficient Markowitz portfolios \cite{markovitz_portfolio_theory}.
		%
		%
		Therefore any solution of the problem 
		\eqref{eq:max_pi_cont} is in particular an efficient Markowitz portfolio.
	\end{remark}
	
	We will need a notation: suppose $\Sigma$ is non-singular; for any $\Lcal \subset \{1,...,K\}, \Lcal \neq \emptyset$ denote by 
	$\Sigma_{\Lcal,\Lcal}^{-1}$ the matrix that, restricted to the indices in $\Lcal$ is the inverse of the $\Lcal \times \Lcal$ minor of $\Sigma$ and zero elsewhere~\footnote{Such a matrix will for instance allow to solve equations of the type $\Sigma x = y$ when both $x$ and $y$ are supported in $\Lcal$; in this case
		$ x = \Sigma_{\Lcal,\Lcal}^{-1} y$.}.  
	
	We give now the main result that shows, under appropriate technical hypothesis, that the output allocation $\pi_t$ will converge to the optimum allocation $\pi^\star$.

	\begin{proposition}
		In the framework above  assume that $\Sigma$ is non-singular. Then~: 
		\begin{enumerate}
			\item maximization problem \eqref{eq:max_pi_cont}  has a unique solution
			$\pi^\star \in \Scal_K $;  \label{item:optimization_unique}
			\item the reward $R_t = R(\pi_t)$ is monotonically increasing;   \label{item:reward_increasing}
			\item 
			The output allocation $(\pi_{t})_{t\ge0}$ in \eqref{eq:def_pit_cont} converges, we denote $\pi^\infty : = \lim_{t\to\infty} \pi_t$; in addition 
			\begin{equation}
				\pi^\infty \in \Scal_K \cap 
				\left\{	
				\one_\Lcal \odot\Sigma^{-1}\mu + \frac{1-\langle \one_\Lcal , \Sigma^{-1}\mu\rangle }{\langle \one_\Lcal ,\Sigma_{\Lcal,\Lcal}^{-1}\one_\Lcal \rangle } \Sigma_{\Lcal,\Lcal}^{-1}  \one_\Lcal,
				\Lcal \subset \{1,...,K\}, \Lcal \neq \emptyset	
				\right\}; \label{eq:pit_limit_set}
			\end{equation} 
			\label{item:convergence_to_some_point}
			\item  \label{item:cv_to_optimal_if_close}
			there exists $c_\epsilon$ depending only on $\Sigma$ and $\mu$ such
			that if $ \| \pi_0- \pi^\star \|_\Sigma \le c_\epsilon$ then $\lim_{t\to\infty}\pi_t =\pi^\star$;
			\item 
			\label{item:cv_to_optimal}
			for general initial value $\pi_0$, not necessarily close to $\pi^\star$,
			if 
			$\pi^\infty \in \intsimplexk$ 
			then 
			$\pi^\star\in \intsimplexk$ and 
			$\lim_{t\to\infty}\pi_t =\pi^\star$.  Moreover, in this case the convergence is exponential i.e. there exists $c_0, c_1 > 0$ such that~: 
			\begin{equation}
				\| \pi_t - \pi^\star\| \le c_0 e^{-c_1 t}, \ \forall t \ge 0.
			\end{equation}
		\end{enumerate}
		
		  \label{prop:convergence}
		 \end{proposition}
	\begin{proof}
		\noindent
		{\bf Proof of step \ref{item:optimization_unique}~:}
		Note that when $\Sigma$ is non singular the maximum in~\eqref{eq:max_pi_cont} is necessarily unique because $\Sigma$ will be strictly positive definite so 
		the maximization problem involves a strictly convex function on the closed convex domain $\Scal_K$.

		Since $\Sigma$ is non-singular, we can assign $\pi^\dagger:=\Sigma^{-1} \mu$. Note that in general $\pi^\dagger$ is not in $\Scal_K$: entries may be negative and their sum is not necessarily equal to $1$. 
		We also introduce the norm $\|x\|^2_\Sigma = \langle \Sigma x,x\rangle$.
		Then 
		\begin{equation}
			R(\pi)=  \frac{1}{2}
			\langle \Sigma \pi^\dagger, \pi^\dagger \rangle - 
			\langle \Sigma \pi - \pi^\dagger, \pi - \pi^\dagger\rangle 
			=  \frac{1}{2} \| \pi^\dagger \|^2_\Sigma -  \frac{1}{2} \| \pi-\pi^\dagger \|^2_\Sigma.
		\end{equation}
		This means that in particular $\pi^\star$ will be the projection of $\pi^\dagger$ on $\Scal_K$ with respect to the norm  $\|\cdot \|^2_\Sigma$.
		
		\noindent
		{\bf Proof of step \ref{item:reward_increasing}~:}
		From \eqref{eq:def_algo_cont} we derive~:
		\begin{equation}
			\frac{d}{dt} R(\pi_t) =
			\frac{d}{dt} R(S(\Hcal(t))) = \left\langle \nabla_H R(S(\Hcal(t))), \frac{d}{d t} \Hcal(t) \right\rangle
			=  \|\nabla_H R(S(\Hcal(t))) \|^2 \ge 0, \label{eq:evolution_r_s_h_cont}
		\end{equation}
		thus $R(\pi_t)$ is increasing.
		
		\noindent
		{\bf Proof of step \ref{item:convergence_to_some_point}~:}
		From the definition of $\pi^\dagger$ we obtain
		$ \nabla_\pi R(\pi)=\Sigma (\pi^\dagger-\pi)$.
		For any column vector $\zeta \in \R^K$ we introduce the  matrix $\mathfrak{H}(\zeta)= diag(\zeta)- \zeta \zeta^T$. Note that $\mathfrak{H}(\zeta)$ acts on a vector $v$ by
		$\mathfrak{H}(\zeta)v = \zeta \odot (v-\bar{v}\one )$ with $\bar{v} = \langle \zeta, v\rangle$. The softmax derivation rule \eqref{eq:softmax_derivation_rule}
		can be written as~:
		$\nabla_H S(H) = \mathfrak{H}(S(H))$. 
		We obtain
		\begin{equation}
			\nabla_H R(S(\Hcal(t))) = \nabla_H \pi_t \nabla_\pi R(\pi_t) 
			= - \mathfrak{H}(\pi_t) \Sigma (\pi_t - \pi^\dagger).
		\end{equation}
		So finally, $\pi_t$ satisfies the following equation
		\begin{equation}
			\frac{d}{dt} \pi_t = \nabla_H \pi_t \frac{d}{dt} \Hcal(t) = - \mathfrak{H}^2(\pi_t) \Sigma (\pi_t - \pi^\dagger). \label{eq:pi_evolution_cont}
		\end{equation}
		In \eqref{eq:pi_evolution_cont} there is no direct dependence of $\Hcal(t)$ but only of  $\pi_t$, so \eqref{eq:pi_evolution_cont} can be considered an autonomous ODE involving $\pi$. This ODE
		leaves invariant  $\intsimplexk$  i.e., if
		$\pi_0 \in \intsimplexk$ then $\pi_t \in \intsimplexk$ $\forall t \ge 0$; to see this it is enough to switch back to the $\Hcal$ formulation and to invoke the uniqueness of the solution. 
		In fact 
		the whole $\Scal_K$ will be invariant for \eqref{eq:pi_evolution_cont}~: for instance direct computations show that if $\pi_t(k)=0$ then
		$\left( \frac{d}{dt}\pi_t \right)(k)=0 $ so  $\pi_t(k)$ will not change sign; in addition 
		the linear constraint $\langle \one,\pi_t \rangle =1$ remains true by continuity. 
		We invoke now LaSalle's invariance principle for the dynamical system \eqref{eq:pi_evolution_cont} set on $\Scal_K$ and Lyapunov function $V(\pi) =-R(\pi)$. We saw from 
		\eqref{eq:evolution_r_s_h_cont} that 
		\begin{equation}
			\dot{V}(\pi) = - \|  \mathfrak{H}(\pi) \Sigma(\pi-\pi^\dagger)  \|^2 \le0, \ \forall \pi \in \Scal_K.
		\end{equation}
		
		Consider now the set
		$E= \{\pi \in \Scal_K : \dot{V}(\pi)=0 \}$. Any $\pi \in E$ will satisfy  
		$ \mathfrak{H}(\pi) \Sigma(\pi-\pi^\dagger) =0$ or equivalently
		$\pi(k)(v_k-\bar{v}\cdot \one)=0$ for all $k\le K$, where $v=\Sigma(\pi-\pi^\dagger)$ and 
		$\bar{v}= \langle \pi, v\rangle$. Denote $\Lcal= \{ k \le K : \pi(k) \neq 0\}$. Previous relation means that $\forall k \in \Lcal~:~ v_k=\bar{v}$, i.e. $\one_\Lcal \odot v = c \cdot \one_\Lcal$ with $c$ a constant.
		Replacing $v$ with its definition we obtain $\one_\Lcal \odot  \Sigma(\pi-\pi^\dagger) = c \cdot \one_\Lcal$ and furthermore 
		$\pi = \one_\Lcal \odot \pi^\dagger + c \Sigma_{\Lcal,\Lcal}^{-1} \one_\Lcal$.
		After taking the scalar product with $\one_\Lcal$ we obtain 
		$c = \frac{1-\langle \one_\Lcal , \pi^\dagger\rangle }{\langle \one_\Lcal ,\Sigma_{\Lcal,\Lcal}^{-1} \one_\Lcal \rangle }$ and therefore
		\begin{equation}
			\pi =  \one_\Lcal \odot\pi^\dagger + \frac{1-\langle \one_\Lcal , \pi^\dagger\rangle }{\langle \one_\Lcal ,\Sigma_{\Lcal,\Lcal}^{-1} \one_\Lcal \rangle } \Sigma_{\Lcal,\Lcal}^{-1}  \one_\Lcal. \label{eq:pi_crit_inv}
		\end{equation}
		This implies that $E$ is discrete with at most $2^K-1$ elements, one for each possible  
		$\Lcal \subset \{1,...,K\}$, $\Lcal \neq \emptyset$.  By LaSalle's principle $\pi_t$ approaches $E$  but since $E$ is discrete  $\pi_t$ will even converge to some point of $E$ denoted $\pi^\infty$.
		
		\noindent
		{\bf Proof of step \ref{item:cv_to_optimal_if_close}~:} by strict convexity, $R(\zeta)< R(\pi^\star)$ for any $\zeta \in E$, $\zeta \neq \pi^\star$. Since the reward is increasing, should $\pi_0$ be close enough to $\pi^\star$ then $R(\pi_0)> R(\zeta), \forall \zeta \in E, \zeta \neq \pi^\star$; since $R(\pi_t)$ is monotonically increasing, $\pi_t$ cannot converge to any such $\zeta$. The only point left to converge is $\pi^\star$. This proves in particular that $\pi^\star \in E$.
		
		\noindent
		{\bf Proof of step \ref{item:cv_to_optimal}~:} 
		Since $\pi^\infty \in \intsimplexk$ 
		the support $\Lcal$ of $\pi^\infty$ is
		$\Lcal = \{ 1,..., K\}$; then
		by the formula  \eqref{eq:pi_crit_inv} 
		\begin{equation}
			\pi^\infty=  \pi^\dagger + \frac{1-\langle \one , \pi^\dagger\rangle }{\langle \one ,\Sigma^{-1} \one \rangle } \Sigma^{-1}  \one. \label{eq:piinf_min}
		\end{equation}
		But the right hand side of \eqref{eq:piinf_min} is the definition of the minimum of 
		$R(x)$ under the sole constraint that $\langle x, \one \rangle=1$. The set of such $x$ is larger than $\Scal_K$ but if the minimum belongs to $\Scal_K$ it will also be the best among elements of $\Scal_K$  so $\pi^\infty=\pi^\star$.
		
		
		We now prove the exponential convergence.
		Since $\mathfrak{H}(\pi_t)\one=0$ for any $\pi_t\in\Scal_K$ we can write~: 
		\begin{equation}
			\mathfrak{H}(\pi_t)\Sigma (\pi_t - \pi^\dagger) 
			=\mathfrak{H}(\pi_t)\Sigma (\pi_t - \pi^\star + c \Sigma^{-1}\one) 
			=\mathfrak{H}(\pi_t)\Sigma (\pi_t - \pi^\star)  + c \mathfrak{H}(\pi_t)\one
			=\mathfrak{H}(\pi_t)\Sigma (\pi_t - \pi^\star),
		\end{equation}
		where $c$ is the constant in \eqref{eq:piinf_min}. 
		Since $\pi_t \to \pi^\star$ and  $\pi^\star \in \intsimplexk$ there exists some $b>0$ small enough and $t_b$ large enough such that 
		$\pi_t(k)\ge b$ for all $k\le K$ and $t\ge t_a$. Let us compute 
		\begin{eqnarray}
			& \ & 
			\frac{d}{dt} \frac{1}{2}
			\left\langle  \Sigma (\pi_t - \pi^\star), \pi_t - \pi^\star\right\rangle
			= \left\langle \Sigma (\pi_t - \pi^\star) , \frac{d}{dt} \pi_t\right\rangle
			=- \left\langle \Sigma (\pi_t - \pi^\star)  ,  \mathfrak{H}^2(\pi_t) \Sigma (\pi_t - \pi^\star) \right\rangle
			\nonumber \\ & \ & 
			= - \|  \mathfrak{H}(\pi_t)\Sigma (\pi_t - \pi^\star) \|^2
			\label{eq:estimation_cv_exp1}\end{eqnarray}
		Denote $v = \Sigma (\pi_t - \pi^\star)$; then~: 
		\begin{equation}
			\|  \mathfrak{H}(\pi_t)\Sigma (\pi_t - \pi^\star) \|^2
			=
			\|  \mathfrak{H}(\pi_t)v\|^2 = \|\pi_t \odot (v - \langle \pi_t,v\rangle) \cdot \one\|^2 \ge 
			b^2 \| v - \langle \pi_t,v\rangle \cdot \one\|^2, \  \forall t \ge t_b. \label{eq:estimation_cv_exp2}
		\end{equation}
		Furthermore, for any vector $v$ the mapping $ \gamma \in \R \mapsto \| v -\gamma \cdot \one\|^2$ is minimized for 
		$\gamma = \frac{\langle \one, v\rangle}{\langle \one, \one\rangle}$ and therefore 
		$ \| v - \langle \pi_t,v\rangle\cdot \one \|^2 \ge  \| v - \frac{\langle \one, v\rangle}{\langle \one, \one\rangle}\cdot \one \|^2$.

		In the compact domain $ \left\{w \in \R^K : \| w \|=1,  \langle w , \one \rangle = 0 \right\}$ the function
		$w \mapsto  \left\| \Sigma w - \frac{\langle \one, \Sigma w \rangle}{\langle \one, \one\rangle}\cdot \one \right\|^2$ has a positive minimum. 
		If this minimum is zero then it is attained for $w^\star$ such that 
		$\Sigma w^\star = \frac{\langle \one, \Sigma w^\star \rangle}{\langle \one, \one\rangle} \one$ thus $ w^\star = \frac{\langle \one, \Sigma w^\star \rangle}{\langle \one, \one\rangle} \Sigma^{-1}\one$; but 
		$0 = \langle \one, w^\star \rangle  = \frac{\langle \one, \Sigma w^\star \rangle}{\langle \one, \one\rangle} \langle \one,  \Sigma^{-1}\one \rangle$. Since  $\Sigma$ is positive definite $\langle \one,  \Sigma^{-1}\one \rangle \neq 0$ and we conclude that $\langle \one, \Sigma w^\star \rangle=0$ which shows that in fact $\Sigma w^\star=0$ thus $w^\star=0$ in contradiction with the requirement 
		that $\|w^\star \| = 1$.
		So, we can conclude that the minimum is not null. 
		Denote it by $m>0$; $m$ only depends on the matrix $\Sigma$. 
		When $\langle w , \one \rangle = 0 $ but $\|w\|$ is not necessarily equal to one the relationship becomes, by proportionality~: 
		$\left\| \Sigma w - \frac{\langle \one, \Sigma w \rangle}{\langle \one, \one\rangle}\cdot \one \right\|^2 \ge m \|w\|^2$. Take now the particular value $w = \pi_t - \pi^\star$, 
		that has indeed $\langle w , \one \rangle = 0$. Recall that $v = \Sigma w$ and, thus 
		$ \left\| v - \frac{\langle \one, v\rangle}{\langle \one, \one\rangle}\cdot \one \right\|^2 \ge 
		m \| \pi_t - \pi^\star \|^2$; 
		since $\Sigma$ is non-singular 
		we obtain finally from all the above considerations, equation 
		\eqref{eq:estimation_cv_exp1} and \eqref{eq:estimation_cv_exp2} that 
		$\frac{d}{dt} \frac{1}{2}
		\left\langle  \Sigma (\pi_t - \pi^\star), \pi_t - \pi^\star\right\rangle \le -b^2 m\|\pi_t - \pi^\star\|^2 \le 
		-c_m \left\langle  \Sigma (\pi_t - \pi^\star), \pi_t - \pi^\star\right\rangle$ for 
		some $c_m >0$ and all $t\ge t_b$. It follows that the norm 
		$\left\langle  \Sigma (\pi_t - \pi^\star), \pi_t - \pi^\star\right\rangle$ converges exponentially to zero and by norm equivalence also does $\| \pi_t - \pi^\star\|^2$.
	\end{proof}
	\begin{remark}
		The hypothesis are mostly technical and can be weakened.
		In particular one can prove that $\pi^\star$ is the only stable critical point for the $\pi_t$ dynamics so (numerically)  $\lim_{t\to\infty} \pi_{t} = \pi^\star$ even  
		without the hypothesis in step~\ref{item:cv_to_optimal}.
	\end{remark}
	
\begin{remark}
Proposition~\ref{prop:convergence} assumes zero transaction costs. In practice, such costs exist but are typically an order of magnitude smaller than the other terms in the ODE; they can therefore be incorporated through a perturbation analysis using standard results such as Gronwall's lemma~\cite{gronwall_lemma}. This does not imply that transaction costs have a negligible impact on the final portfolio value, only that the allocation can remain close to optimal. The total portfolio value still depends on turnover and is empirically analyzed in the next section (see Figure~\ref{fig:ir_ka_fee2_turnover}).

Finally, even if the theoretical result does not go beyond Gaussian return distributions, we will see  empirically that the algorithm performs satisfactory even for distributions with non-negligible kurtosis ('heavy tails'), see statistics in Table~\ref{table:nyse_o_couples}.

In full rigor, a proof of the Proposition~\ref{prop:convergence} in presence of non-zero transaction costs and non-Gaussian distributions would be interesting but we leave it for future work.
\end{remark}
	
\section{Numerical results and discussion}\label{sec:numerical}
	
	To implement the Onflow algorithm \ref{algo:onflow}, one has to 
	select the dataset, the fee level $\xi$, specify the parameter $\tau$ and solve ODE \eqref{eq:ode_formula}. The ODE resolution is very robust with respect to the method used; in practice we employed the \texttt{odeint} routine from the \texttt{SciPy} Python package, see \cite{2020SciPy-NMeth}. No special option was used and all other parameters are left to their default settings.
	
	For the numerical tests we use the "Old NYSE" database, a benchmark from the literature   listing the prices of $36$ stocks quoted on the New York Stock Exchange from $1965$ to $1987$ ($5651$ daily prices i.e., $T=5650$), see \cite{cover91,Helmbold98,kalai2002efficient} and \cite[the "nyse\_o.csv" file]{marigold_github_up}.
	We take pairs of assets as described in 
	Table \ref{table:nyse_o_couples} which reproduces the presentation from 
	\cite[p. 122]{dochow_proposed_2016}. All distributions display positive excess kurtosis ('heavy tails') significant to p-value $\le 0.001$ (see also Remark~\ref{rem:kurtosis}).

	 \begin{table}[h]
		\caption{Descriptions of the pairs tested in section~\ref{sec:numerical}. The 'correlation' row refers to the correlation between price relatives $f^1_t$ and $f^2_t$ (not between absolute prices $S^1_t$ and $S^2_t$).
		}\label{table:nyse_o_couples}%
		\begin{tabular}{@{}c|c|c|c|c@{}}
			\toprule
			No. & 1 & 2 & 3 & 4 \\
			\midrule
			Asset names & \begin{tabular}[c]{@{}c@{}}Commercial Metals\\Kin Ark\end{tabular} & \begin{tabular}[c]{@{}c@{}}Iroquois\\Kin Ark\end{tabular} & \begin{tabular}[c]{@{}c@{}}Coca Cola\\IBM\end{tabular} & \begin{tabular}[c]{@{}c@{}}Commercial Metals\\Meicco\end{tabular} \\
			\midrule
			Correlation & 0.064 & 0.041 & 0.388 & 0.067 \\
			\midrule
			Individual  & 52.02& 8.92 & 13.36 &52.02\\
			performances &4.13 &4.13 & 12.21 & 22.92\\
			\midrule

			{Individual  }& {40.32\%}& {54.45\% }& {22.24\% }&{40.32\%}\\
			{volatilities }& {79.04\%}& {79.04\%}& {21.23\% }& {48.93\%}\\
			\midrule
			
			{Individual  }& {0.0222}& {0.0142 }& {0.0193 }&{0.0222}\\
			{excess kurtosis }&{0.0112 }&{0.0112 }& {0.0103 }& {0.0316}\\
			\midrule

			 Description & \begin{tabular}[c]{@{}c@{}}Volatile,  stagnant\\uncorrelated\end{tabular} & \begin{tabular}[c]{@{}c@{}}Volatile\\uncorrelated\end{tabular} & \begin{tabular}[c]{@{}c@{}}Non-volatile\\highly correlated\end{tabular} & Volatile \\
			\bottomrule
		\end{tabular}
	\end{table}
	
\begin{figure}[htpb!]
		\centering
		\includegraphics[width=.95\linewidth]{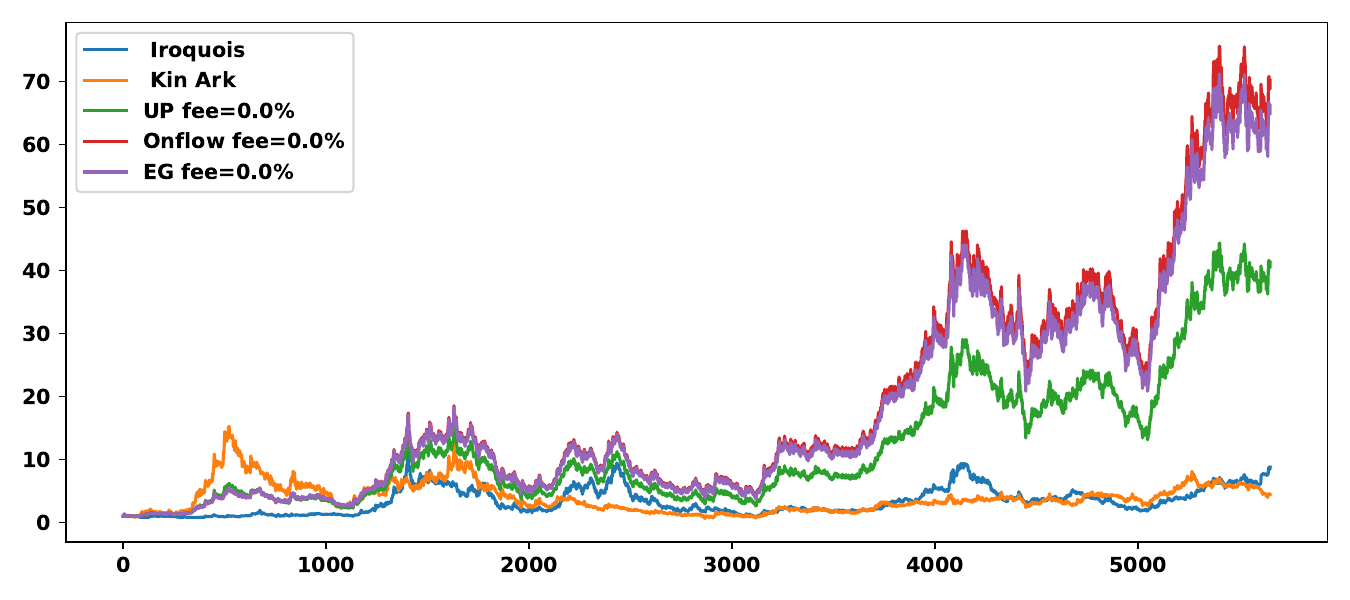}
		\caption{Results for the pair 'Iroquois' -- 'Kin Ark', fee level=$0\%$: evolution of the UP, EG and Onflow portfolios. EG and Onflow perform similarly, better than UP which in turn is better than the  individual assets. 
		}\label{fig:ir_ka_fee0}
	\end{figure}

\begin{figure}[htpb!]
		\centering
		\includegraphics[width=.95\linewidth]{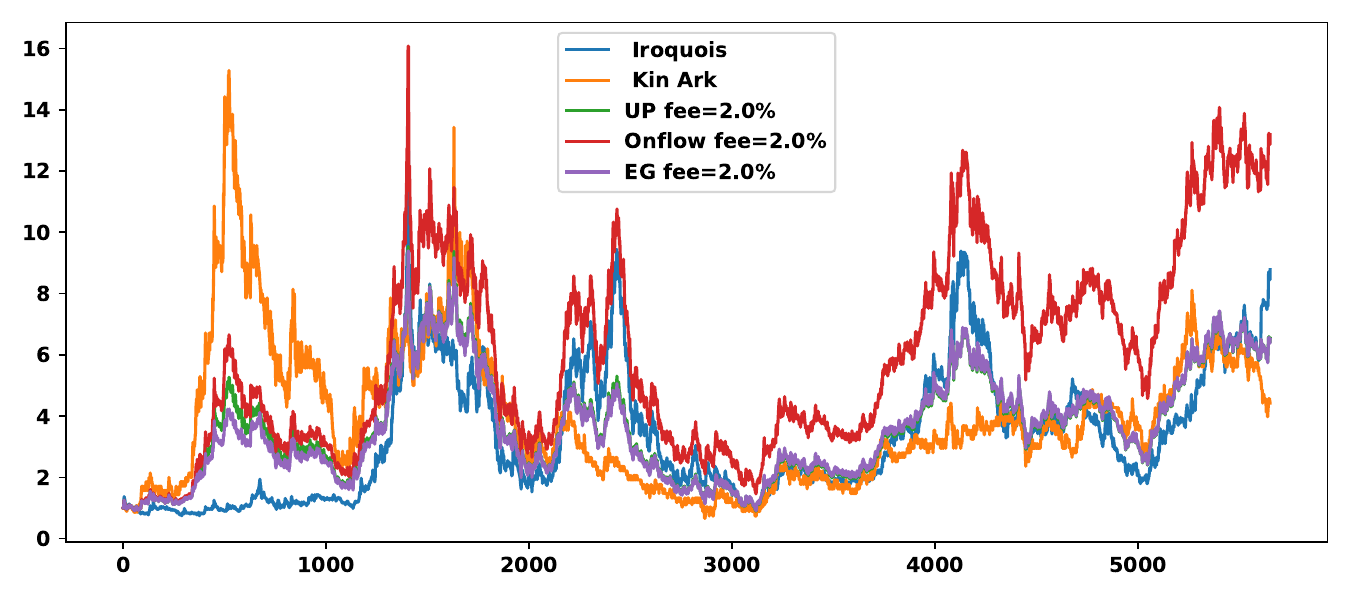}
		
		\includegraphics[width=.95\linewidth]{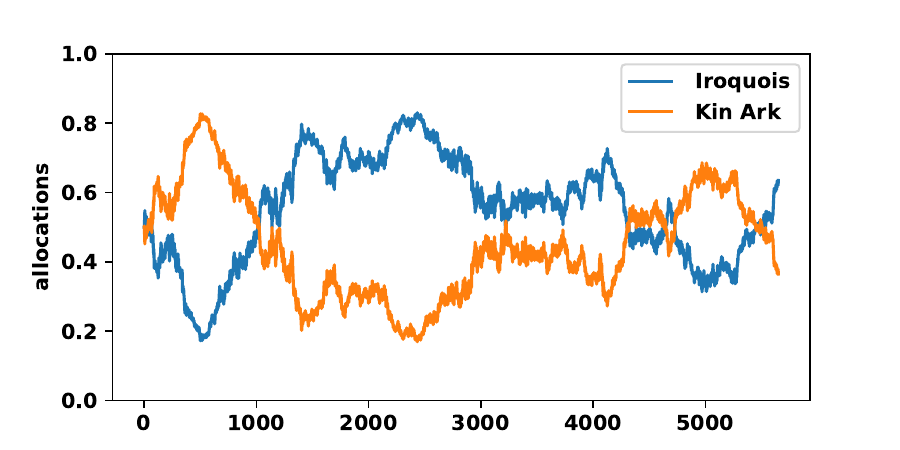}
		\caption{Results for the pair 'Iroquois' -- 'Kin Ark', fee level=$2\%$.  {\bf Top~:} evolution of the UP, EG and Onflow portfolio. With this fee level only the Onflow portfolio performs better than the individual assets. 
			{\bf Bottom~:} the allocations of the Onflow portfolio.
		}\label{fig:ir_ka_fee2}
	\end{figure}

\begin{figure}[htpb!]
		\centering
		\includegraphics[width=.95\linewidth]{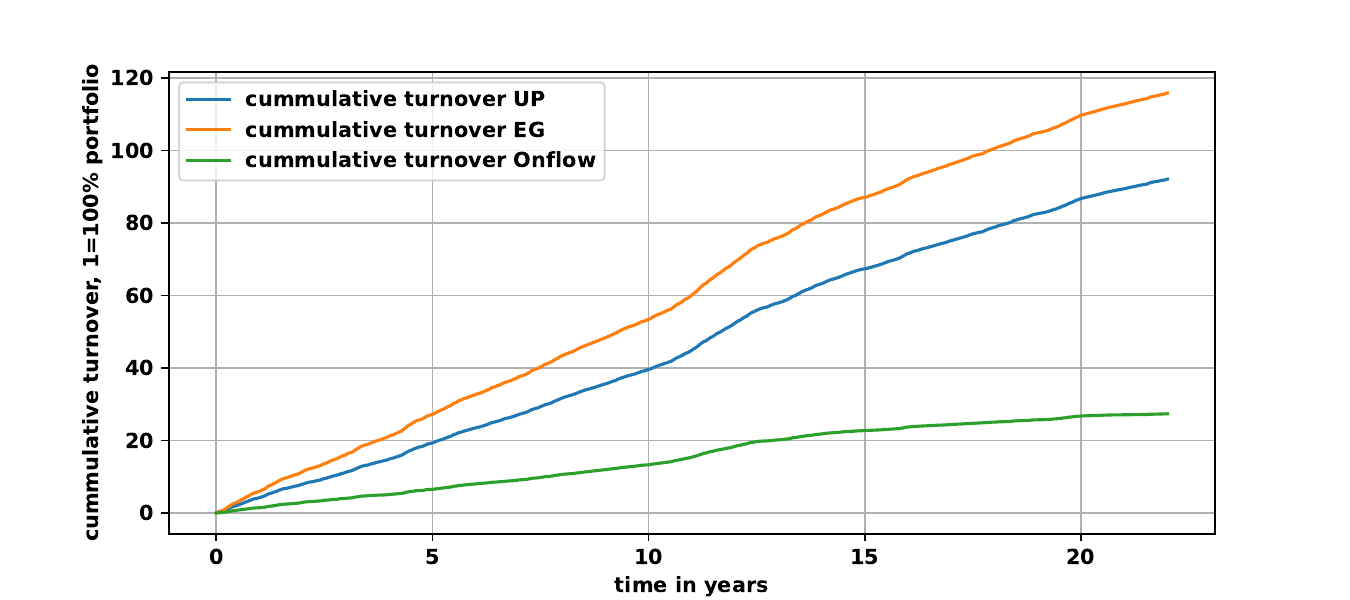}
		
		\caption{Cumulative turnover for the pair 'Iroquois' -- 'Kin Ark', fee level=$2\%$:  Units are set such that a value of $1$ corresponds to a $100\%$ portfolio turnover. For instance the total turnover over the whole period for UP is around $90$ times the portfolio value (not to be mistaken with $90\%$!).}\label{fig:ir_ka_fee2_turnover}
	\end{figure}

\begin{figure}[htpb!]
		\centering
		\includegraphics[width=.95\linewidth]{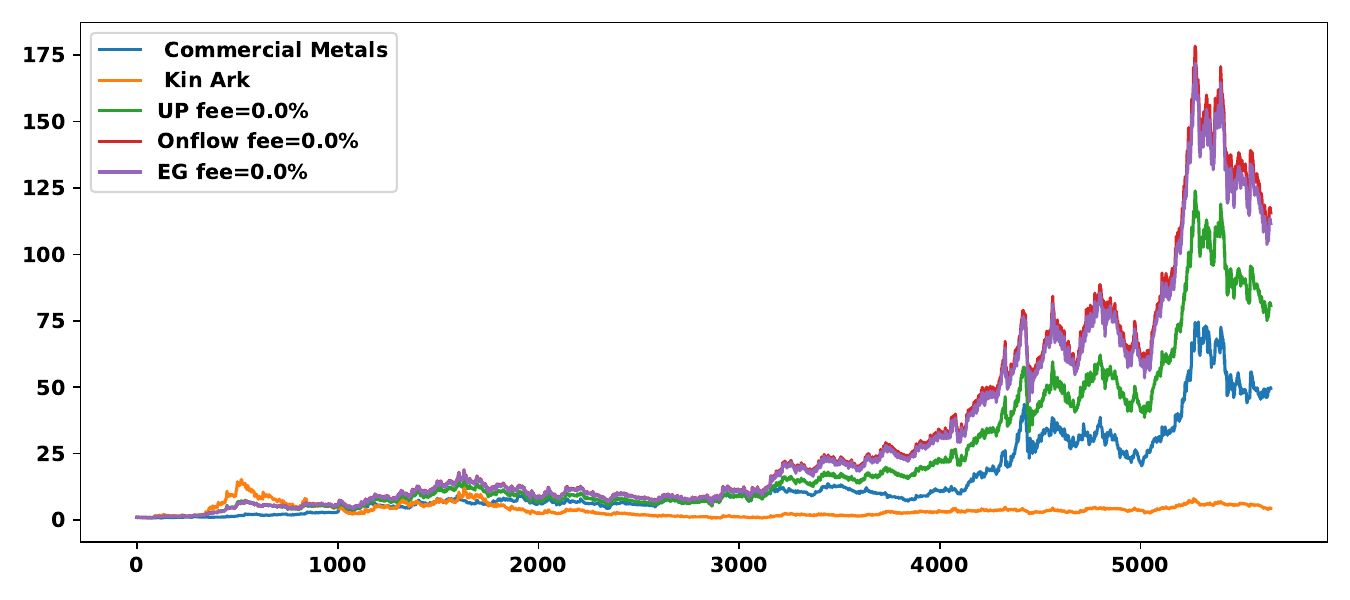}
		\caption{Same results as in Figure \ref{fig:ir_ka_fee0} for the pair 'Commercial Metals' -- 'Kin Ark', $\xi=0\%$. 
		}\label{fig:com_ka_fee0}
	\end{figure}

\begin{figure}[htpb!]
		\centering
		\includegraphics[width=.95\linewidth]{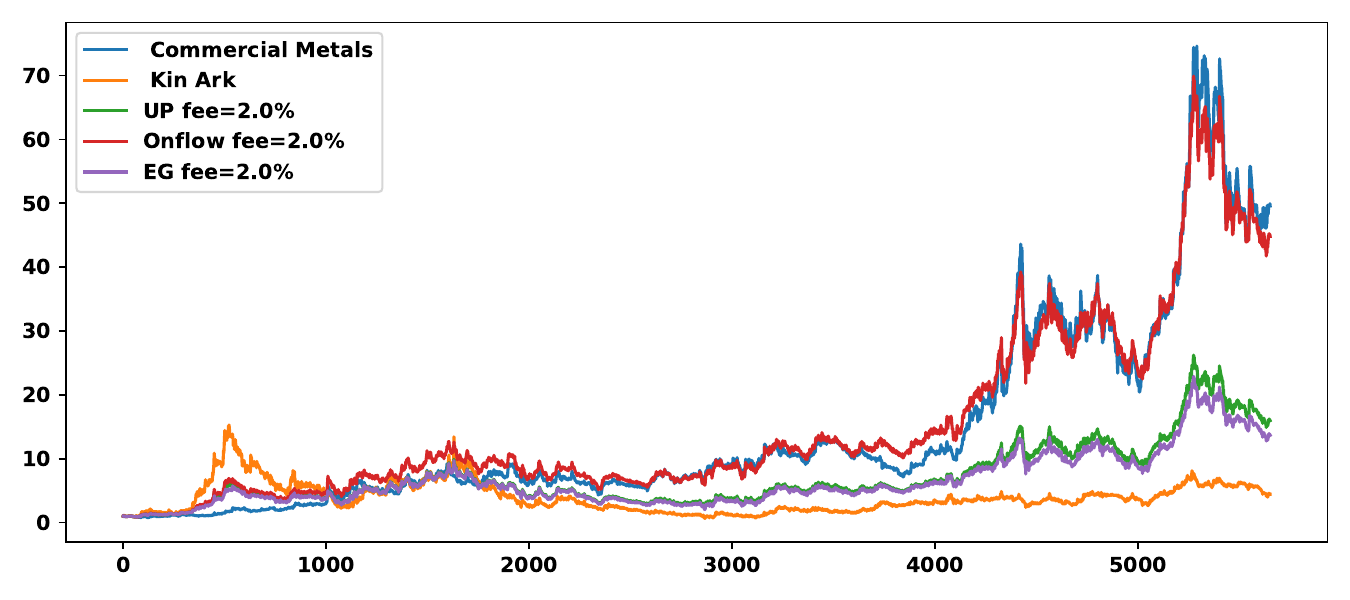}
		
		\includegraphics[width=.95\linewidth]{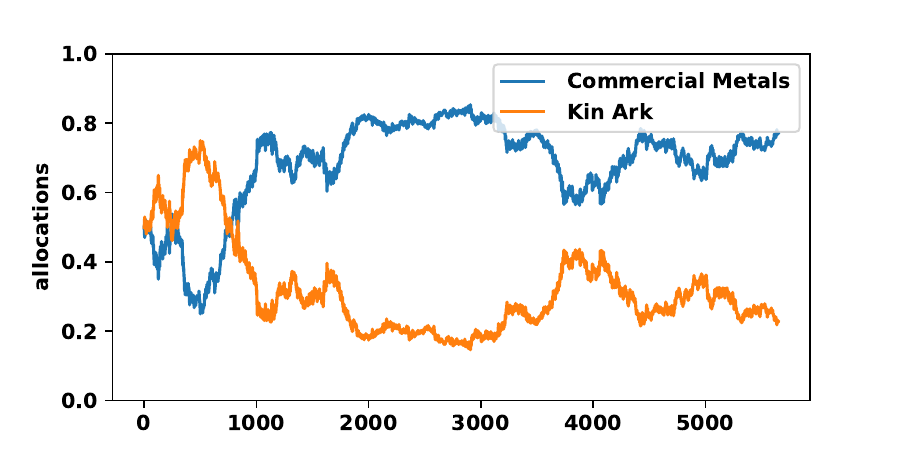}
		\caption{Same results as in Figure \ref{fig:ir_ka_fee2} for the pair 'Commercial Metals' -- 'Kin Ark', $\xi=2\%$. 
		}\label{fig:com_ka_fee2}
	\end{figure}

\begin{figure}[htpb!]
		\centering
		\includegraphics[width=.95\linewidth]{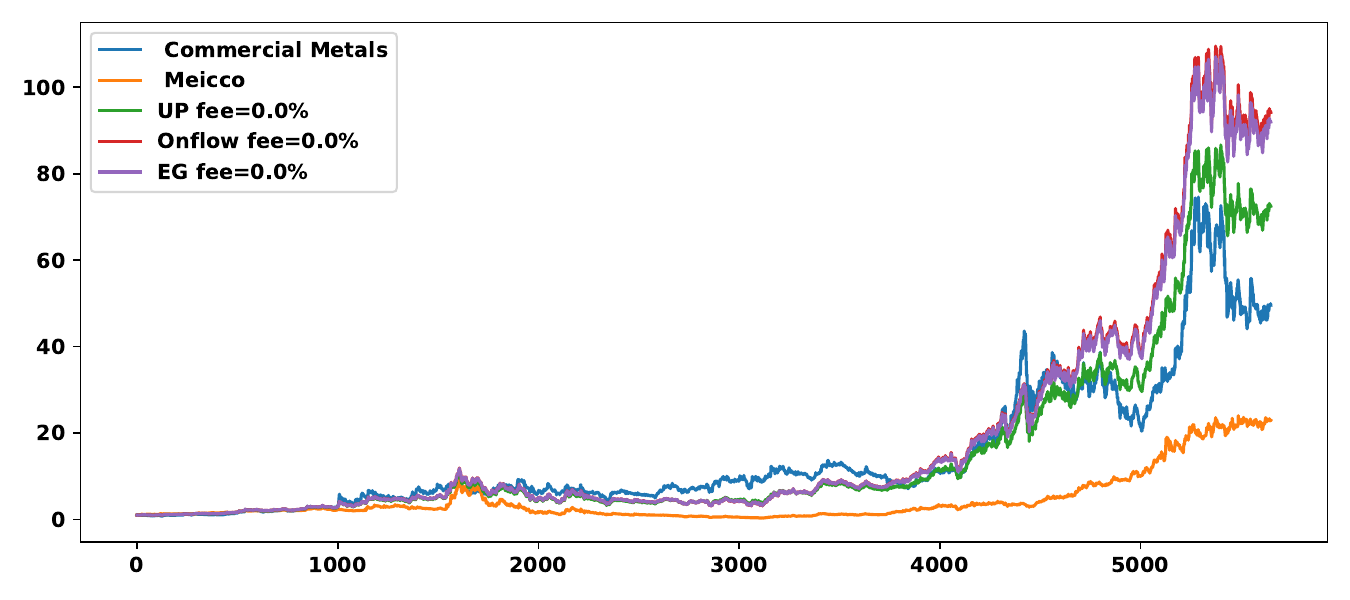}
		\caption{Same results as in Figure \ref{fig:ir_ka_fee0} for the pair 'Commercial Metals' -- 'Meicco', $\xi=0\%$. 
		}\label{fig:com_mei_fee0}
	\end{figure}

\begin{figure}[htpb!]
		\centering
		\includegraphics[width=.95\linewidth]{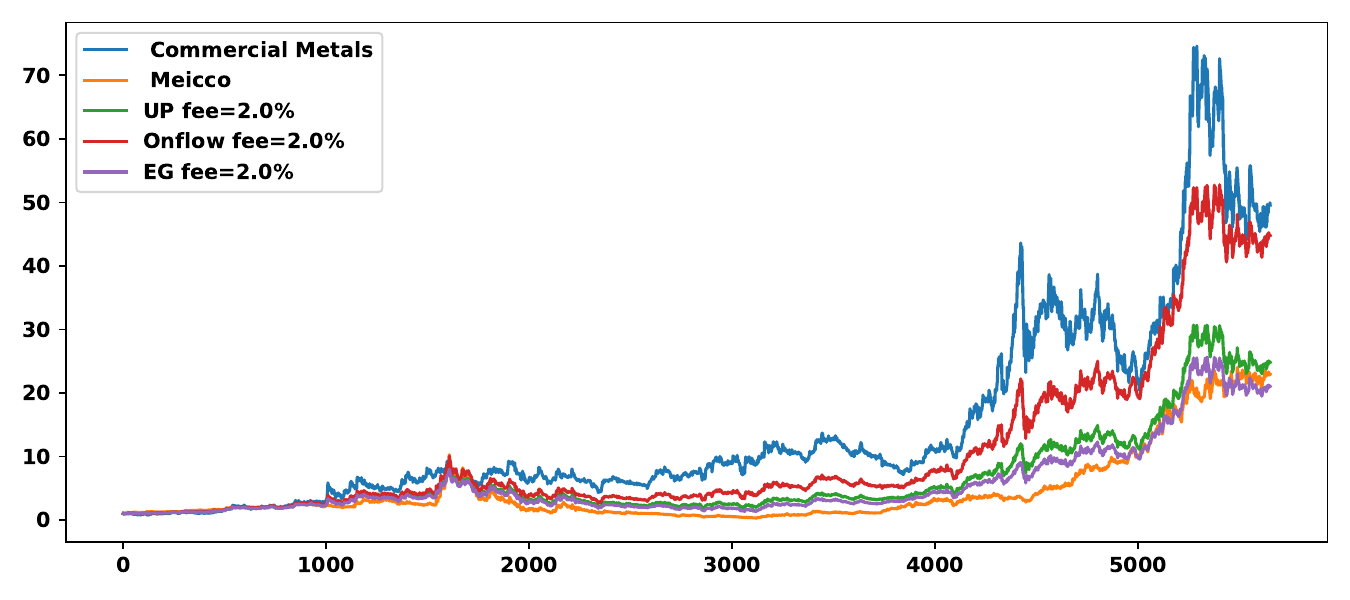}
		
		\includegraphics[width=.95\linewidth]{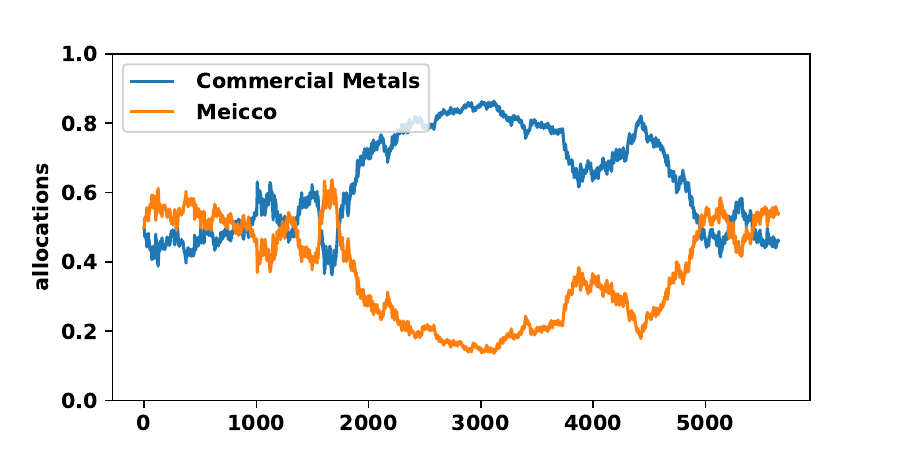}
		\caption{Same results as in Figure \ref{fig:ir_ka_fee2} for the pair  'Commercial Metals' -- 'Meicco', $\xi=2\%$. 
		}\label{fig:com_mei_fee2}
	\end{figure}

\begin{figure}[htpb!]
		\centering
		\includegraphics[width=.95\linewidth]{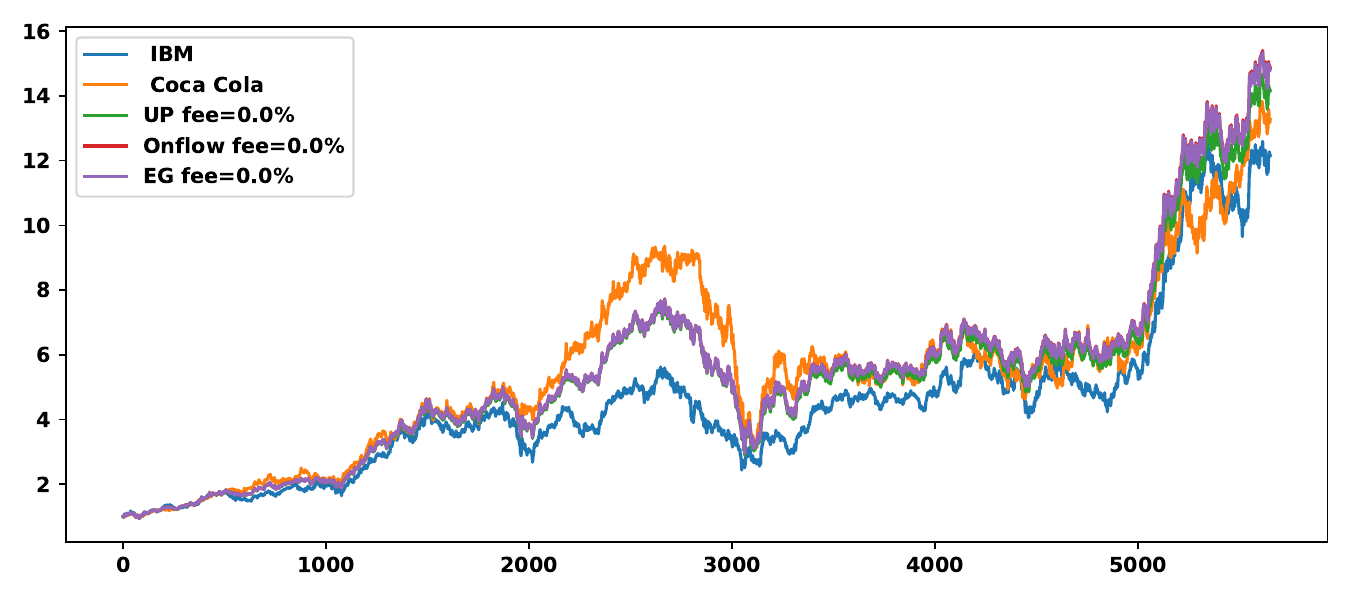}
		\caption{Same results as in Figure \ref{fig:ir_ka_fee0} for the pair 'IBM' -- 'Coca Cola', $\xi=0\%$. 
		}\label{fig:ibm_cc_fee0}
	\end{figure}

\begin{figure}[htpb!]
		\centering
		\includegraphics[width=.95\linewidth]{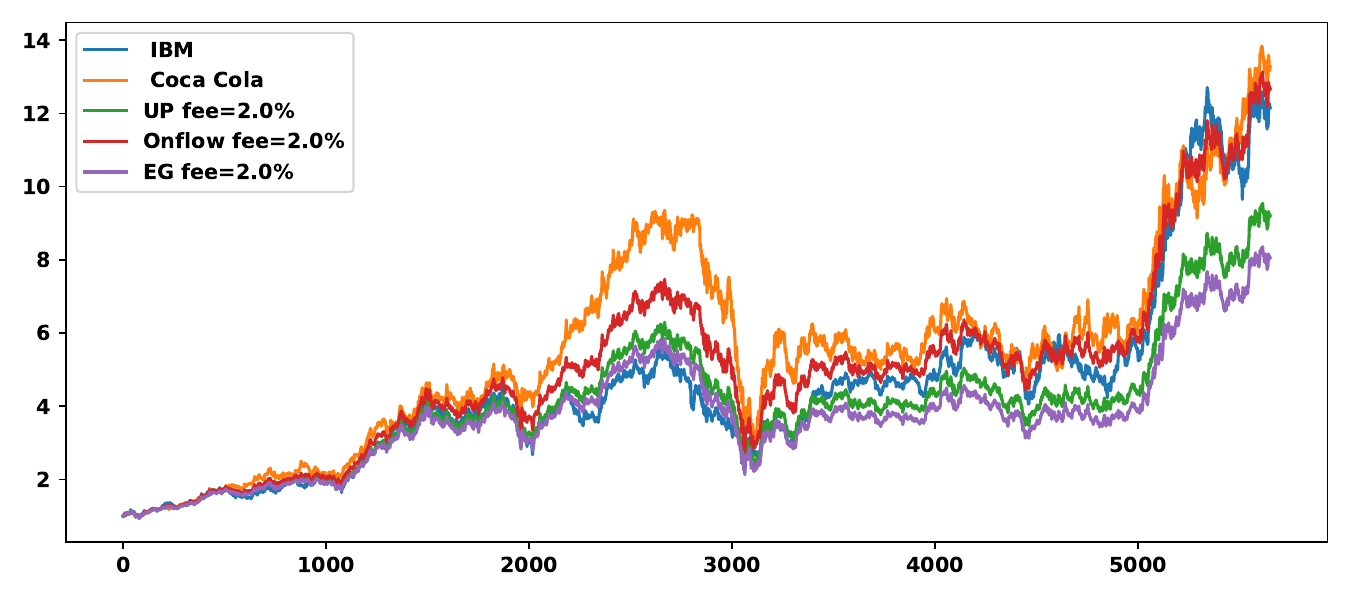}
		
		\includegraphics[width=.95\linewidth]{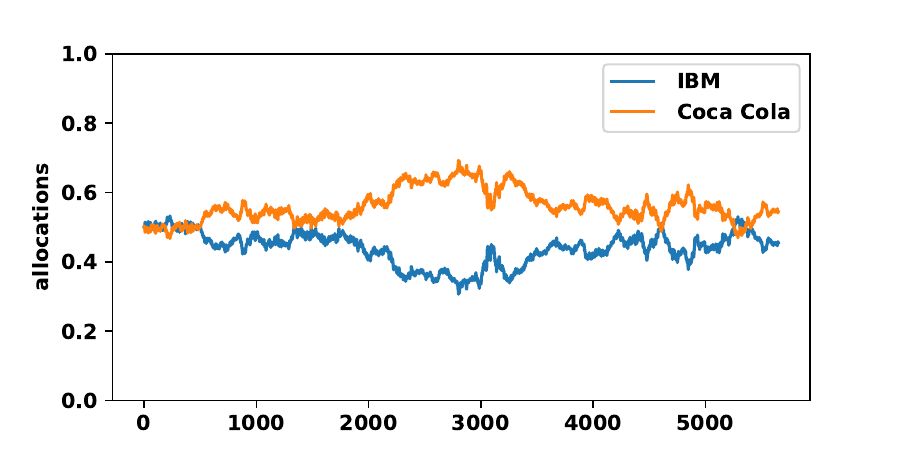}
		\caption{Same results as in Figure \ref{fig:ir_ka_fee2} for the pair 'IBM' -- 'Coca Cola', $\xi=2\%$. 
		}\label{fig:ibm_cc_fee2}
	\end{figure}

	In all situations we plot the results for two fee values $\xi=0$ and $\xi=2\%$ and the time evolution of the value of several portfolios~: the individual assets, the Cover Universal portfolio labeled 'UP', the \cite{Helmbold98} portfolio (label 'EG') with parameter $\eta$  set to $\eta=0.05$ as in the reference and the Onflow portfolio, parameter $\tau$ set to 
	$0.05$ when $\xi=0$ and $\tau=1$ when $\xi=2\%$.\footnote{The hyper-parameter $\tau$ was chosen after some empirical tests; it  depends only on the transaction costs parameter $\xi$ and not on the assets tested. In general we recommend to set  $\tau$ as an increasing function of $\xi$;
			unfortunately, at this moment, a principled justification for selecting $\tau$ for a given value of $\xi$ is not yet available but may be object of future work. Nevertheless, empirical back-testing usually helps obtaining effective values.
	 } \footnote{
		We focus on strategies related to reinforcement learning. Our method may not necessarily perform better than other procedures in all circumstances; note in particular that not all share the same optimization criteria (VaR, Sharpe ratio and so on).}
	Note that a transaction fee of $2\%$ is usually very difficult to handle and the performance of  most of the known algorithms collapses in this case.
	We now review the results presented in Figures \ref{fig:ir_ka_fee0}-\ref{fig:ibm_cc_fee2}. 
	
	A pair known to provide good performance (cf. \cite{cover91}) is 'Iroquois' and 'Kin Ark'  (Figures \ref{fig:ir_ka_fee0} and \ref{fig:ir_ka_fee2}).
	The individual stocks increase by a factor of $8.92$ and $4.13$ respectively, while UP obtains around $40$ times the initial wealth. Even more, EG and Onflow manage to obtain around $70$ times the initial wealth, which is a substantial improvement over UP (and individual stocks). Even if Onflow is slightly better than EG, the difference does not seem to be substantial. On the other hand, when the fee level $\xi$ increases to $2\%$ 
	the performance of all the portfolios except Onflow degrade to the point of not being superior to that of simple buy-and-hold strategies on individual stocks. This result is consistent with  the literature, which highlights the severe impact of the transaction costs on dynamic portfolio strategies. Here, the Onflow parameter $\tau$ was set to $1$.

	The cumulative turnover (often called "rotation rate" in fund prospectus)
	is plotted in Figure~\ref{fig:ir_ka_fee2_turnover}; when $\xi=0$ 
	the daily portfolio turnover  $\sum_t |\pi_{t+1}-\pi_{t+}|$ (mean relative transaction volume) is around $2\%$ for all strategies  UP, EG and Onflow ;  when $\xi=2\%$ UP and EG keep the turnover at the same level while Onflow reduces it to $0.5\%$. This explains the performance of  Onflow in this case. Note that a level of daily turnover of $2\%$ corresponds to over 
	$500\%$ annual turnover while  $0.5\%$ means about $125\%$ annually. Over the whole period of $22$ years, UP and EG have a turnover 
	of around $100$ times the portfolio value while Onflow has a total turnover  $\sim 25$.
	
	Our second test is the pair 'Commercial Metals' -- 'Kin Ark' (Figures \ref{fig:com_ka_fee0}-\ref{fig:com_ka_fee2}). The same general conclusions hold here, with the performance of 
	individual stocks not exceeding $50$ times initial wealth, 
	Cover UP being above this at around $80$ while EG and Onflow are above UP at around $110$ when $\xi=0$. When $\xi=2\%$ the performance deteriorates~: UP and EG decrease to $\sim 15$ while Onflow manages to retain cca. $50$ times initial wealth. In this case the reason is simple~: in hindsight the 'Commercial Metals' has a very impressive performance over the period and the best thing to do it is to passively follow it. This is what the Onflow algorithm manages to do as one can see in the bottom plot of Figure \ref{fig:com_ka_fee2} which shows that past
	the time $1000$ the allocation of "Commercial Metals" is always superior to that of 'Kin Ark' and goes often as high as $80\%$ of the overall portfolio.
	
	The results for the pair 'Commercial Metals' -- 'Meicco', are presented in Figures
	\ref{fig:com_mei_fee0} and \ref{fig:com_mei_fee2}. As before, the impressive performance of the 'Commercial Metals' stock does not allow for much improvement, with the Onflow algorithm remaining competitive even when fees are taken into account.

	Finally, we consider a situation where dynamic portfolios do not work well, the pair 'IBM' -- 'Coca Cola'. Without transaction costs all portfolios are comparable to individual assets. However 
	at $2\%$ fee level  UP and EG are not as good as the individual stocks while Onflow manages to obtain comparable results.
	
\begin{remark}
Since the results have not been computed on log-normal assets,  
the empirical findings suggest that variations in kurtosis do not necessarily bias the algorithm in the search for optimal allocations.
\label{rem:kurtosis}
\end{remark} 
	
	\subsection{{Further tests on high volatility assets}}
	\label{sec:high_vol}

	\begin{figure}[htpb!]
		\centering
		\includegraphics[width=.75\linewidth]{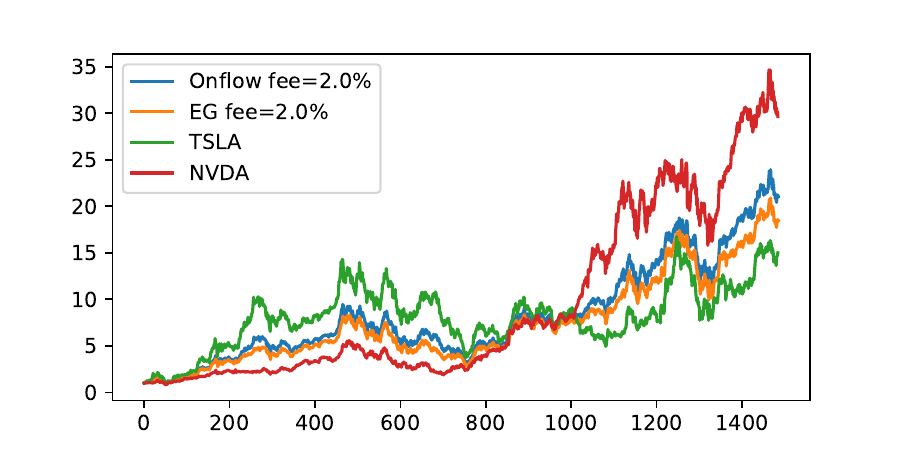}
		
		\includegraphics[width=.75\linewidth]{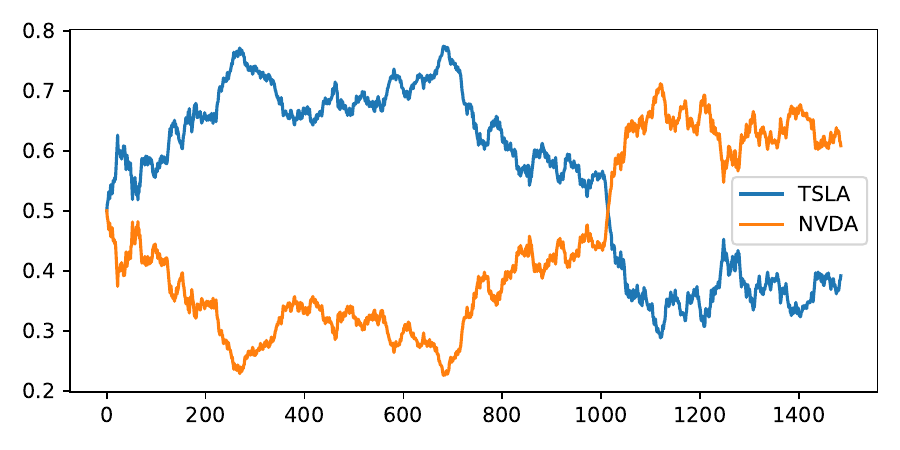}
		\caption{{Results for high-volatility, high capitalization stocks in \Cref{sec:high_vol}
				, here Nvidia and Tesla; same conventions as before (top: performance, bottom allocation), $\xi=2\%$, $\tau=1$. 
		}}\label{fig:nvda_tsla_fee2tau1}
	\end{figure}

	\begin{figure}[htpb!]
		\centering
		\includegraphics[width=.75\linewidth]{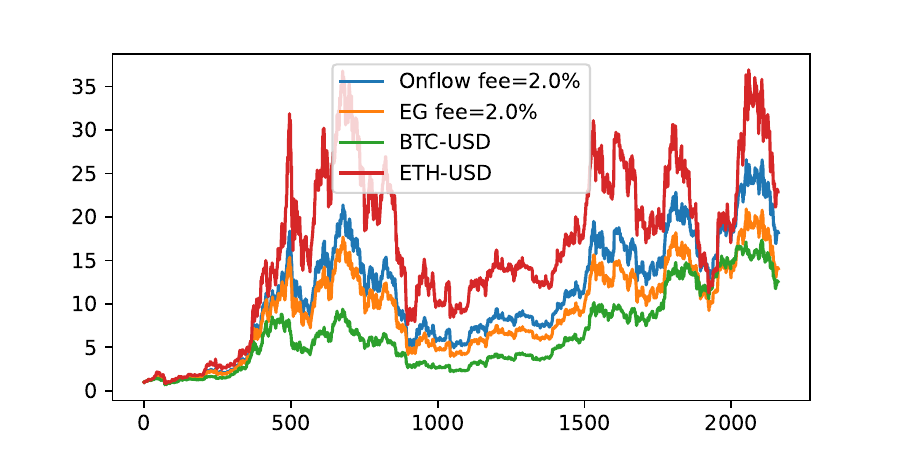}
		
		\includegraphics[width=.75\linewidth]{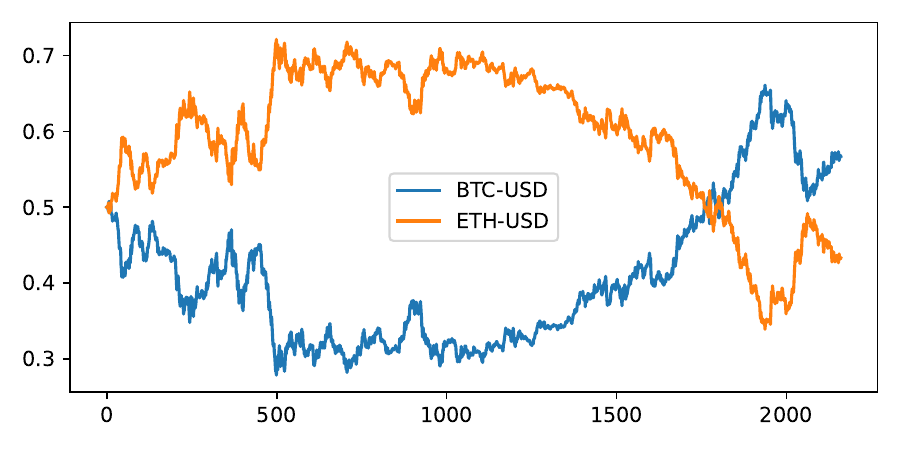}
		\caption{Results for high-volatility cryptocurrencies in 
					\Cref{sec:high_vol}.  Top plot: performance, bottom plot: allocation), $\xi=2\%$, $\tau=1$. 
		}\label{fig:btc_eth_fee2tau1}
	\end{figure}
	
To check the algorithm's performance in different market structures 
		we also considered examples with high volatility. To this end we took couples of assets 
		with large market capitalization and also high volatility for the period 2020-2025; more specifically we considered two cryptocurrencies : the Bitcoin (BTC-USD ticker)
		and Ethereum (ETH-USD) and two impactful tech stocks  Nvidia (NVDA), Tesla (TSLA). 
The dataset is constructed with end of day quotes for the period Jan 1st 2020 to Nov 28th 2025. 

	The results for the tech stocks are in \Cref{fig:nvda_tsla_fee2tau1}; we also calculated the Sharpe ratios: 'Onflow': 1.21, 'EG': 1.19,  'TSLA':0.99, 'NVDA': 1.32. Of course, here NVDA is performing exceptionally over the period and it is difficult to outperform it. However our algorithm is doing a good job at identifying the periods where TSLA is performing better (up to day 1000) and then switching to NVDA afterwards at it quintuples its value during the last two years.
	
	{At the fee level of $2\%$, the Onflow performance degrades moderately from $22.32$ to $21.04$ with a mean daily turnover of $0.2\%$, while the EG is goes from $27.67$
		to $18.46$ due to the mean daily turnover of $1.36\%$.
	}
	
	{The results for the two cryptocurrencies are in \Cref{fig:btc_eth_fee2tau1}; the Sharpe ratio for 'Onflow' is $0.86$, for 'EG' $0.82$, for 'BTC-USD' $0.83$ and for 'ETH-USD' $0.87$. The 'Onflow' performance without transaction fees is 
		$18.67$ and diminishes to $18.19$ for a daily turnover of 
		$0.06\%$ while for the 'EG' strategy it goes from $19.86$
		to $14.04$ for a daily mean turnover of 
		$0.8\%$. 
		The same qualitative remarks are also true here, the 'Onflow' being able to track the relative performances in the allocations. By the end of the period, the 'Onflow' strategy takes the bet to under-allocate ETH-USD despite a higher momentarily performance because of a rapid and worrisome decay in the most recent months. In such a high volatility market this is probably a protective move, and only the future can tell if the bet will turn out to be justified.
	}

	 \subsection{More than two assets: Sharpe ratio results}
	\label{sec:sharpe_results}
	
\begin{figure}[htpb!]
		\centering
		\includegraphics[width=.95\linewidth]{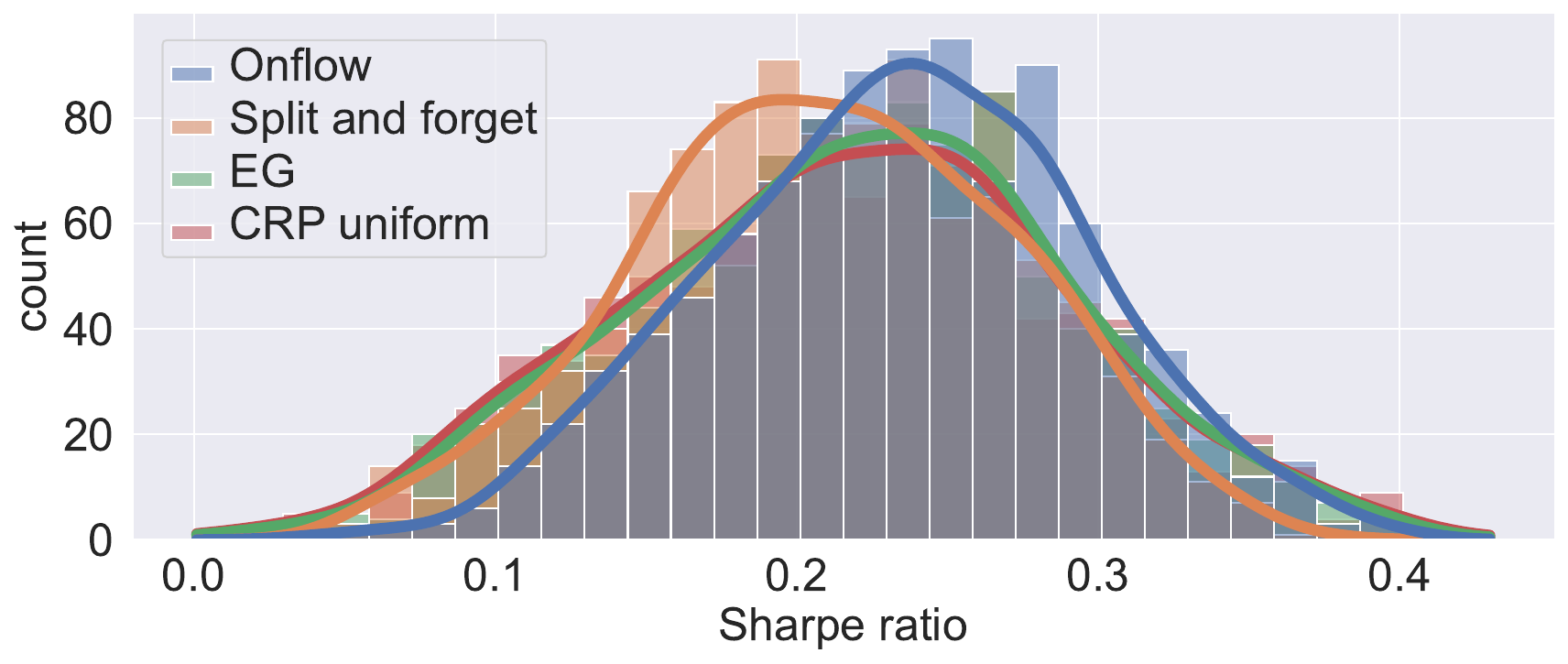}
		\caption{Sharpe Ratio of four strategies as described in Section~\ref{sec:sharpe_results}. Median values of the Sharpe ratios are as follows: 
			Onflow: $0.237$, 
			Split and forget: $0.209$,
			EG                 : $0.221$
			CRP uniform        : $0.218$.		
		}\label{fig:sharpe10}
	\end{figure}

	We test now situations with more than two assets. We consider the same dataset as above but we take at random $1000$ samples each consisting of $10$ assets from the original universe. For each sample we record the Sharpe ratio of four strategies:  {\it Onflow} (ours) the {\it EG} from \cite{Helmbold98} with their parameters, the {\it uniform Constant Rebalanced Portfolio} (CRP)\footnote{Uniform CRP allocates evenly between all assets at each time.} and the {\it Split and Forget} strategy which invests same amount in each asset at initiation and never touches the portfolio again. We take a moderate amount of transaction fee $\xi=1\%$ and set $\tau=2.5$ (empirically optimized value). For each strategy we compute the Sharpe ratio as the mean return minus the risk free rate divided by the standard deviation of the returns. The risk-free return over the period covered by the dataset was estimated from historical records (the Fed Funds) to be $8\%$.
	The results are given in Figure~\ref{fig:sharpe10}. Our strategy does not {\bf always} improve over the other three but a standard Wilcoxon test reveals that the Sharpe ratio of 'Onflow' is superior to each of the other strategies  
	and this difference is statistically significant (p-value  $< 10^{-3}$).
	
	\subsection{{Numerical efficiency}}
	
\begin{figure}[ht]
		\centering
		\includegraphics[width=0.8\textwidth]{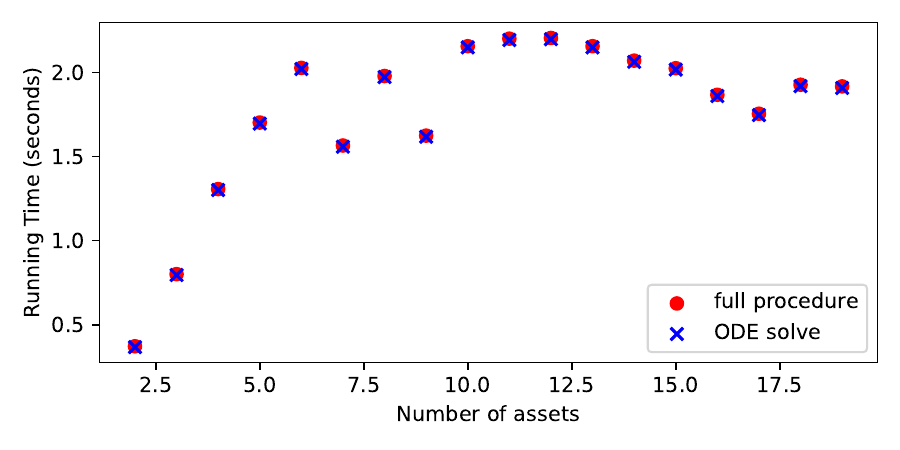}
		\caption{{Computational efficiency analysis: ODE resolution time versus number of assets. The runtime growth rate stabilizes for portfolios over $10$ assets, indicating good scalability. We took a dataset with $252$ days, $\tau=2.5$ and $\xi=2\%$ as in Section~\ref{sec:sharpe_results}. We plot the total run time of the procedure (for the $252$ days) and the total time spent solving ODEs, which is almost the same. 
		}}
		\label{fig:scalability}
	\end{figure}

	{To address the scalability requirement for portfolios with more than $10$ assets, we conducted an additional computational efficiency analysis focusing on ODE solution time and memory usage. As shown in Figure~\ref{fig:scalability}, the runtime initially increases with the number of assets but then stabilizes for larger portfolios (over $10$ assets), demonstrating that the proposed algorithm scales efficiently. We plot both total running time of the algorithm  and the ODE running time to confirm that this is the most expensive part of the computation. The actual running time depends on the hardware and other specifics but the behavior illustrated here appears to be generic.
	}

	\subsection{{Comparison with other low-turnover approaches}}
	\label{sec:compare_w_kirby_et_al}
	
	\begin{figure}[htpb!]
		\centering
		\includegraphics[width=.75\linewidth]{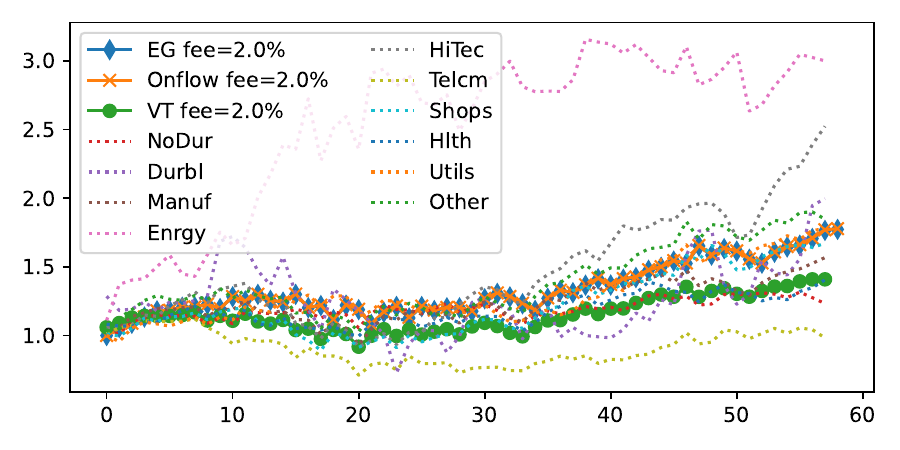}
		
		\includegraphics[width=.75\linewidth]{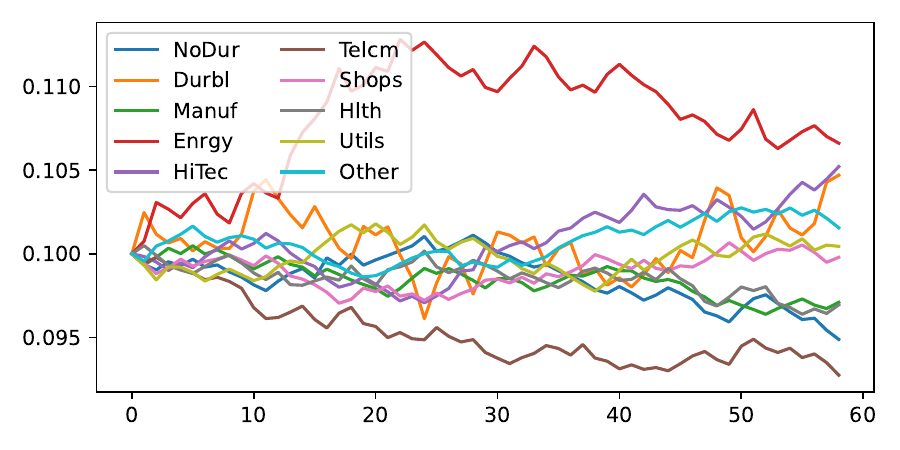}
		\caption{{Results for Volatility Timing strategy from \cite{kirby_low_turnover_12}, see 	\Cref{sec:compare_w_kirby_et_al}; top plot: performance, bottom plot: allocation), $\xi=2\%$, $\tau=1$, monthly data. 
		}}\label{fig:10ind_VT_Kirby_fee2tau1}
	\end{figure}

	\begin{table}
		\caption{{Portfolio Performance Metrics for\Cref{sec:compare_w_kirby_et_al}.}}
		\begin{tabular}{l|cccc}
			\toprule
			{Sector or}& {Performance }& {Sharpe  }& {Average }& {Performance  }\\
			{strategy}&  &  {Ratio }&  {Turnover (\%) }&  {After Fees }\\
			\midrule
			{NoDur }& {1.23 }& {1.83 }& {0.00 }& {1.23 }\\
			{Durbl }& {2.00 }& {1.93 }& {0.00 }& {2.00 }\\
			{Manuf }& {1.57 }& {2.83 }& {0.00 }& {1.57 }\\
			{Enrgy }& {3.00 }& {4.18 }& {0.00 }& {3.00 }\\
			{HiTec }& {2.52 }& {4.55 }& {0.00 }& {2.52 }\\
			{Telcm }& {0.98 }& {0.25 }& {0.00 }& {0.98 }\\
			{Shops }& {1.67 }& {3.11 }& {0.00 }& {1.67 }\\
			{Hlth }& {1.41 }& {2.07 }& {0.00 }& {1.41 }\\
			{Utils }& {1.77 }& {3.79 }& {0.00 }& {1.77 }\\
			{Other }& {1.85 }& {3.54 }& {0.00 }& {1.85 }\\
			\hline \hline
			{Volatility Timing }& {1.59 }& {2.34 }& {10.68 }& {1.41 }\\
			{EG }& {1.84 }& {3.99 }& {3.21 }& {1.77 }\\
			{Onflow }& {1.84 }& {4.01 }& {2.87 }& {1.78 }\\
			\bottomrule
		\end{tabular}
		\label{tab:perf_10Ind}
	\end{table}

To further compare our strategy with established benchmarks from the literature we considered the approach of	\cite{kirby_low_turnover_12} 
that develop methods of mean-variance portfolio selection characterized by low turnover. They prove that 
such procedures outperform naive diversification even in the presence of  transaction costs up to $0.5\%$. As they do, we tested 
the 10 Fama-French (FF) industry portfolios for the period 2020-2025 (monthly data) and employed the 'Volatility Timing' strategy\footnote{We used a Python version of the \texttt{olpsR::alg\_VT} R 'olps' package implementation \cite{olpsR}.}. The results are given in	\Cref{fig:10ind_VT_Kirby_fee2tau1}	and	\Cref{tab:perf_10Ind}; it is seen that, although the Onflow porfolio does not beat the best industrial sector, it has a competitive Sharpe ratio; we believe that the under-performance in terms of final value is also due to the data being somehow scarce (monthly). With respect to other strategies the turnover is very low. Of course, we expect that the Volatility Timing strategy can be tailored to output better results by adjusting for instance the $\eta$ parameter (with the notation from   \cite{kirby_low_turnover_12}) but we took the default value $\eta=1$ here.

	 \section{Conclusion}\label{sec:conclusion}
	We introduce in this paper OnFlow, an online portfolio allocation algorithm. It works without any assumption on the statistics of the asset price time series by repeatedly adjusting the portfolio allocation according to new market data in a reinforcement learning style. Onflow uses a softmax representation of the allocation and solves during each time step a gradient flow evolution equation that can be implemented through a simple ODE; this gradient flow also contains terms to minimize the transaction costs. 
	
	For the case of log-normal continuous time evolution and assuming that true gradients can be used, we show theoretically under some technical assumptions that the procedure will converge to the optimum allocation.
	
	The empirical performance of the procedure was tested on some standard benchmarks with satisfactory results. When compared to classic strategies such as the Universal Portfolio of Cover or the EG algorithm from \cite{Helmbold98} it provides a comparable (even slightly better) performance when transactions fees are zero and performs generally significantly better when severe transactions fees of up to $2\%$ are considered (a level that previous algorithms did not treat very well). 
	
	Some limitations of this study could be the object of future work : 
	add the possibility of short positions (see also \Cref{rem:short}), compare with recent stock prediction machine learning (LSTM networks) algorithms such as 
	\cite{FISCHER18_lstm,Nelson17_lstm_price_prediction,Bhandari22} 
	or test for other markets and different periods.

	\bibliography{refs}
\end{document}